\documentclass[%
 reprint,
 amsmath,amssymb,
 aps,
]{revtex4-2}

\usepackage{graphicx}% Include figure files
\usepackage{dcolumn}% Align table columns on decimal point
\usepackage{bm}% bold math
\usepackage{multirow}
\usepackage{caption}
\usepackage{subcaption}
\usepackage{makecell}
\usepackage{xcolor}
\usepackage[hang]{footmisc}
\usepackage{soul}
\usepackage{tikz}
\usepackage{hyperref}% add hypertext capabilities
\begin{document}

\preprint{APS/23Na3Hed}

\title{Indirect Measurement of the $^{23}\textbf{Na}(p,\gamma)^{24}$Mg Direct Capture Reaction Rate via ($^3$He,d) Spectroscopy}% Force line breaks with \\


\author{Kaixin Song}
\author{Richard Longland}%
\author{Caleb Marshall}%

 \altaffiliation[Present address: ]{Department of Physics, University of North Carolina at Chapel Hill, Chapel Hill, North Carolina, 27599-3255, USA
}
\author{Kiana Setoodehnia}
 \altaffiliation[Present address: ]{Department of Physics, Duke University, Durham, North Carolina, 27708-0305, USA
}
\author{Federico Portillo Chaves}%
\author{Axel Fraud}%
\affiliation{Department of Physics, North Carolina State University, Raleigh, NC 27695, USA}%
\affiliation{Triangle Universities Nuclear Laboratory, Durham, NC 27708, USA}%

% \author{Ann Author}
%  \altaffiliation[Also at ]{Physics Department, XYZ University.}%Lines break automatically or can be forced with \\
%  \email{Second.Author@institution.edu}
% \affiliation{%
%  Authors' institution and/or address\\
%  This line break forced with \textbackslash\textbackslash
% }%

\date{\today}% It is always \today, today,

\begin{abstract}
The cross section of the $^{23}\text{Na}(p,\gamma)^{24}\text{Mg}$ reaction is dominated by direct capture at low energies relevant for stellar burning. Such cross sections can be constrained using spectroscopic factors($C^2S$) or asymptotic normalization coefficients(ANCs) from transfer reactions. In this work, the $^{23}\text{Na}(^3\text{He},d)^{24}\text{Mg}$ reaction was measured at $E_{lab}=21$ MeV to extract spectroscopic factors for $^{24}\text{Mg}$ states with excitation energies in $E_x=7\sim12~$MeV using the Enge split-pole spectrograph at the Triangle Universities Nuclear Laboratory. A new non-resonant astrophysical S factor and the direct capture reaction rate for the $^{23}\text{Na}(p,\gamma)$ reaction are calculated and presented based on this measurement. The new rate at $T<0.04$ GK is 43$\%$ smaller than in previous studies. Rigorous treatments of uncertainties are presented using a Bayesian Markov Chain Monte Carlo (MCMC) method. Sources of uncertainties for computing the direct capture cross section are also discussed in detail.
% \begin{description}
% Secondary publications and information retrieval purposes.
% You may use the \texttt{description} environment to structure your abstract;
% use the optional argument of the \verb+\item+ command to give the category of each item. 
% \end{description}
\end{abstract}

\maketitle

%\tableofcontents

\section{Introduction}\label{sec:Intro}
Globular clusters are collections of tens of thousands of gravitationally-bound stars. They are among the building blocks of our knowledge of the universe \cite{RaffaeleGC}. A common chemical signature of globular clusters is the anti-correlation between oxygen and sodium abundances, observed in Red Giant Branch (RGB) stars across many clusters \cite{Gratton2012}. For decades, the sodium producing and destroying reactions in the NeNa cycle have been measured to explain the observed abundances. At astrophysically low temperature (T$<$50MK), the direct capture component of the sodium-destroying $^{23}\text{Na}(p,\gamma)^{24}\text{Mg}$ reaction determines the sodium abundance observed in RGB stars. The Coulomb barrier reduces the cross section significantly, and thus it is technically challenging to directly measure the direct capture cross sections at low energies. Experimentally, those cross sections can be determined from the spectroscopic factors, $C^2S$, which can be extracted from proton transfer studies~\cite{SF_DC}.

\begin{figure}[ht]
\includegraphics[width=90mm]{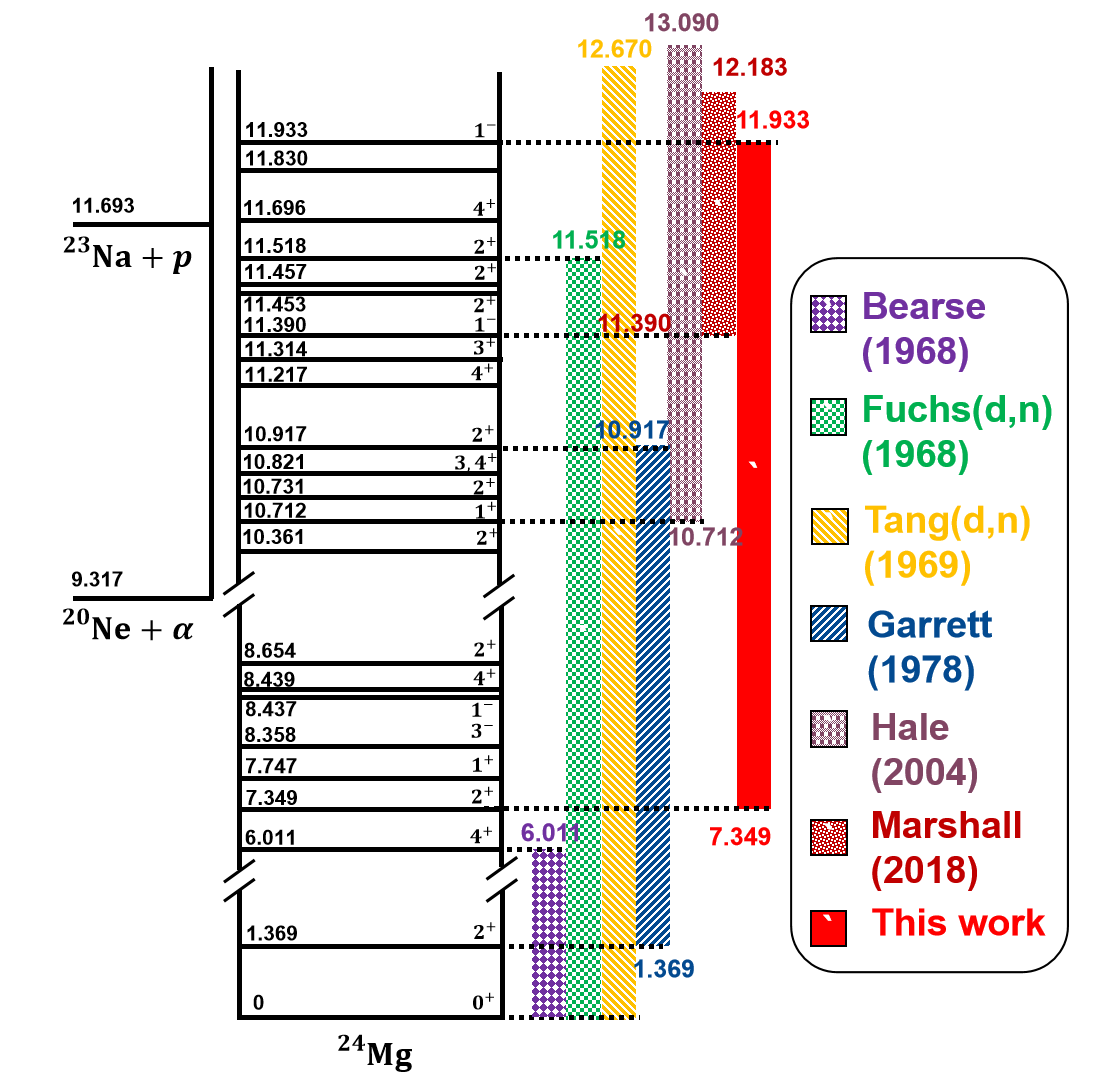}% Here is how to import EPS art  
\caption{\label{fig:Erange} Energy range (in MeV) covered by previous $^{23}$Na$+p$ spectroscopic factor studies.}
\end{figure}

A number of proton transfer studies have been performed \cite{Bearse,Fuchs,Tang,Garrett78,Hale04,Marshall20}. Fig.\ref{fig:Erange} shows the energy ranges measured in those studies. All studies were $(^3\text{He},d)$ measurements except for the $(d,n)$ studies of Fuchs \textit{et al.}~\cite{Fuchs} and Tang \textit{et al.}~\cite{Tang}. Those studies both suffered from poor energy resolution and unknown level properties. Thus, they included much fewer states than the later work. Bearse \textit{et al.}~\cite{Bearse} measured $C^2S$ values only from the ground state to the 6.011 MeV state. $C^2S$ values from the Garrett \textit{et al.}~\cite{Garrett78} measurement have been commonly adopted in recent studies of the $^{23}\text{Na}+p$ direct capture cross sections. Levels below 11 MeV were comprehensively included by Ref.~\cite{Bearse,Garrett78}. Their resultant $C^2S$ values carried large uncertainties due to: (i). Beam and target (composition, degradation, thickness, etc.) effects that introduced uncertainties to the absolute reaction cross sections. (ii). Higher-energy states that were not measured. (iii). Ambiguities from unresolved adjacent states or transitions with mixed angular momenta. A recent calculation from Boeltzig \textit{et al.}~\cite{Boeltzig22} showed that the 10.7 MeV state contributes the most ($\sim$23\%) to the total direct capture cross section using the $C^2S$ from Ref.~\cite{Garrett78}. Later, the 10.7 MeV state was resolved as a doublet of a $J^\pi=1^+$ state and a $J^\pi=2^+$ state by Hale \textit{et al.}~\cite{Hale04}. Hale \textit{et al.} provided data with better resolution, but did not include levels below 10.7 MeV. Boeltzig \textit{et al.}~\cite{Boeltzig22} also pointed out that the spectroscopic factors from Garrett \textit{et al.}~\cite{Garrett78} are the main data set constraining the direct capture cross section calculations. Their experimental results suggest that the spectroscopic factors of Garrett \textit{et al.} may have been twice as large, which implied that additional proton transfer measurements is needed to constrain the uncertainties.

In this work, re-analysis of the data from a recent $^{23}\text{Na}(^3\text{He},d)$ measurement was performed. The experiment and data were first presented by Marshall \textit{et al.} in Refs.~\cite{Marshall2021} and \cite{Marshall23}. Those studies analyzed the $^{24}$Mg states above the proton separation energy $S_p=11.69$ MeV and addressed their impacts on the resonant reaction rates reaction rates. In this work, the $C^2S$ for levels near and below 11 MeV were extracted from the same experimental data, with several states in overlap to check for the consistency. The uncertainties for the $C^2S$ were propagated with a Markov Chain Monte Carlo (MCMC) method developed by Marshall \textit{et al.}~\cite{Marshall20,Marshall23}. An up-to-date $^{23}\text{Na}+p$ direct capture (DC) reaction rate was then calculated and presented using the updated $C^2S$ values from this work with statistically rigorous uncertainties.

In section II, our experimental details are presented. Section III shows our data analysis including the energy calibration, the Distorted Wave Born Approximation (DWBA) and Bayesian MCMC method. Section IV discusses the discrepancies among previous $C^2S$ studies and presents our DC reaction rate. Section V contains our conclusions and final recommendations.

\section{Experiment}\label{sec:exp}

The experiment was performed at Triangle Universities Nuclear Laboratory. Details about the experiment are presented in Ref.\cite{Marshall23}. Briefly, a $^3$He$^{++}$ beam of 21 MeV energy was accelerated by the FN tandem accelerator and delivered to bombard a $\approx 70 \mu g/cm^2$ NaBr target with 15-25 $\mu g/cm^2$ $^{nat}C$ backing in a cylindrical target chamber. Typical beam currents were 100-200 enA, as measured using a suppressed beam stop behind the target. Target degradation was monitored using elastically scattered $^3$He with a silicon detector telescope mounted at 45$^{\circ}$ relative to the incoming beam inside the target chamber. The solid angle for the silicon detector was measured to be $\Omega_{\text{Si}}=4.23(4)$ msr. Downstream the target, the outgoing reaction products were momentum-analyzed by an Enge split-pole spectrograph. They were focused on the focal plane and detected by a detector consisting of two position-sensitive gas avalanche counters, a $\Delta E$ proportionality counter and a residual $E$ scintillator. More details about the detector can be found in Ref.~\cite{FPDetector}. The solid angle of the spectrograph was fixed at $\Omega_{\text{SPS}}=1.00(4)$ msr during the experiment. 

In earlier studies such as Ref.\cite{Garrett78} and Ref.\cite{Hale04}, spectroscopic factors, $C^2S$, were typically extracted from the experimentally determined absolute differential cross sections of the transfer reaction. In this work, both $^3$He elastic scattering and proton transfer cross sections were measured by the focal plane detection system relative to the elastic scattering cross sections measured by a Si $\Delta$E-E telescope placed at 45 degree in the target chamber. This removes any systematic uncertainties arising from the target composition and fluctuations in beam current. As such, the experiment involved two data collections: First, to obtain the transfer reaction data used for DWBA analysis, deuteron spectra from the $(^3\text{He},d)$ reaction were collected for $\theta_{\text{lab}}=5^{\circ}$ to 21$^{\circ}$ in 2$^{\circ}$ steps. Next, to deduce the optical model parameters required for the DWBA analysis, $^3$He spectra were collected from $\theta_{\text{lab}} = 15^{\circ}$ to $55^{\circ}$ in $5^{\circ}$ steps and at $59^{\circ}$. The spectrograph magnetic field was set to $1.14-1.15$T for the transfer runs and to $0.75-0.80$T for the elastic scattering runs. The position of the detector was moved to the calibrated focal plane position at each angle based on kinematics of the reaction. 

In this experiment, $^{24}$Mg states in the energy range of $7\sim13$ MeV are observed. The experimental data were first analyzed by Marshall \textit{et al.}~\cite{Marshall23} for the proton unbound and near-threshold states ($E_x\geq11.39$ MeV). %The near-threshold and unbound states ($E_x\geq11.39$ MeV) were analyzed and discussed by Marshall \textit{et al.} in Ref.~\cite{Marshall23} for their resonance contributions to the reaction rates. 
This work focuses on the bound states ($E_x<11.69$ MeV) and their direct capture contributions.

% {\color{red}{Comment from Richard}}

\begin{figure*}[ht]
\includegraphics[width=170mm]{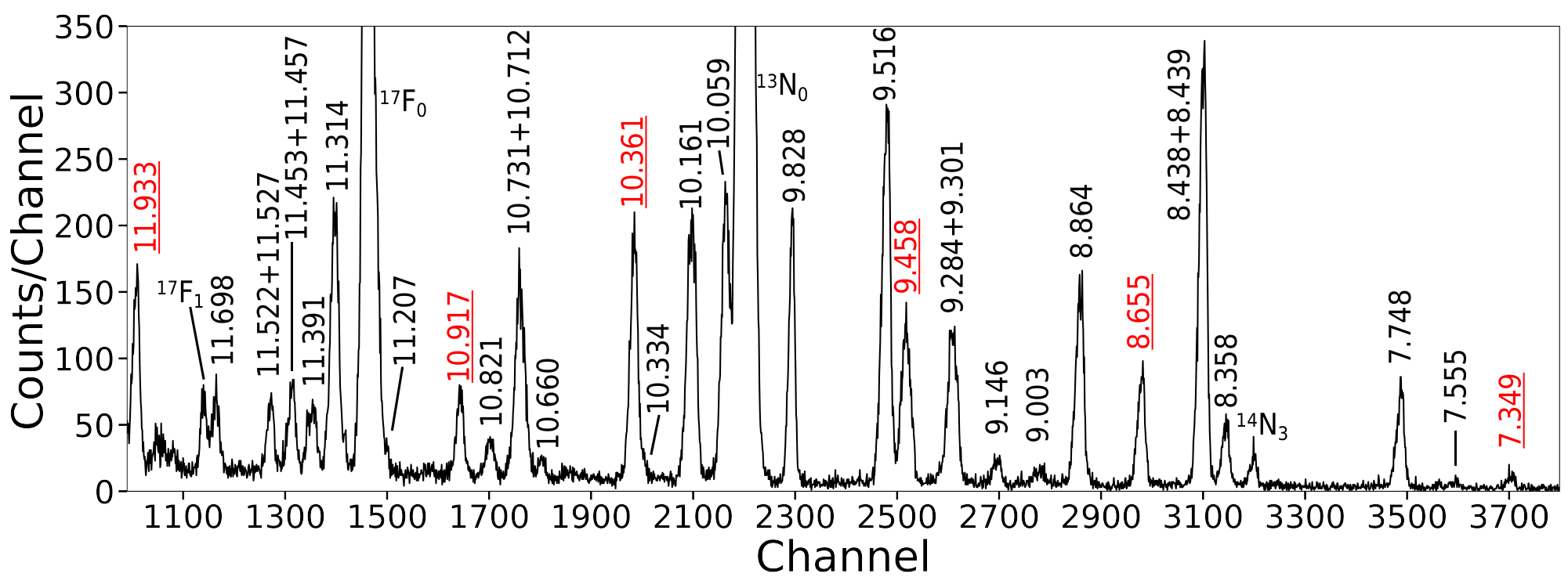}% Here is how to import EPS art
\caption{\label{fig:spec} Deuteron spectrum collected by the focal plane detector at $\theta_{lab}=13^\circ$. Peaks corresponding to $^{24}$Mg states are labeled with energy values taken from Ref.~\cite{ENSDFNOV22}. The energies are in MeV. The red and underlined states were used for the energy calibration. $^{17}$F$_0, ^{17}$F$_1, ^{13}$N$_0, ^{14}$N$_3$ denote contaminant peaks from the ground and first excited states of $^{17}$F, the ground state of $^{13}$N and the 3rd excited state of $^{14}$N, respectively.}
\end{figure*}

\section{Data Analysis}\label{sec:Data}

\subsection{Excitation energies}\label{subsec:Ecal}
A deuteron spectrum collected at $\theta_{\text{lab}}=13^{\circ}$ is shown in Fig.\ref{fig:spec}. Note that only the energy range of interest in this work is shown in the plot. A complete spectrum can be found in Ref.~\cite{Marshall23}. The labeled energies for $^{24}$Mg states are cited from the adopted levels in the Evaluated Nuclear Structure Data File(ENSDF) in Nov. 2022 \cite{ENSDFNOV22}. To extract peak positions and areas, individual peaks were fitted with a Gaussian and a linear background. The energy was calibrated at each angle using a 3rd-order polynomial model and uncertainties determined using a Bayesian MCMC method. The calibrated energies deduced from the angular data are compared with those reported in Ref.~\cite{Marshall23} for consistency and are found to be in general agreement. The averaging and uncertainty propagation were evaluated using the methods from Ref.~\cite{Averaging_SchmellingM_1995} to account for possible correlations. A total of 6 well-resolved states were chosen as calibration points. They are marked with underlines in Fig. \ref{fig:spec}. The uncertainties, including both statistical and systematic contributions, were approximately $\pm2$ keV for most of our measured energies. For states that were weakly populated or resolved from doublets, the uncertainty can be $\pm3$ keV or higher.

% \begin{figure}[ht]
% \includegraphics[width=90mm]{Fig/Energies.png}% Here is how to import EPS art
% \caption{\label{fig:energies} Excitation energies and uncertainties derived in this work(red) relative to the ENSDF values(black). Calibration energies are plotted with red circles. Some of the large error bars of ENSDF energies in the plot were combined from adjacent levels corresponding to unresolved doublets in our data. }
% \end{figure}

\subsection{DWBA analysis}\label{subsec:DWBA}

In order to extract spectroscopic factors for bound states in $^{24}$Mg, a reaction model describing both elastic scattering and the proton stripping reaction is required. For this work, the reaction code FRESCO \cite{FRESCO} was used to perform zero-range DWBA calculations. Details about the theoretical mechanism can be found in Ref.~\cite{Thompson_nuc}. The optical model parameters are from the Woods-Saxon(WS) potential:
\begin{align}
&V(R)=-\frac{V_0}{1+\text{exp}(\frac{r-R}{a})}
\label{eq:WS},
\end{align}
where $V_0$, $r$ and $a$ are the depth, radius and diffuseness of the potential. $R$ is the radius of the nucleus: $R=r_0A^{1/3}$, where $r_0=1.25$ fm, $A$ is the mass number of the nucleus. The optical model uses linear combinations of real and imaginary Woods-Saxon potential to describe the volume, surface and spin-orbit interactions between two nuclei.
Three interactions were considered in this work: (i).$^{23}$Na+$^3$He as entrance channel, (ii).$^{24}$Mg+$d$ as exit channel and (iii).$^{23}$Na+$p$ system as a bound state formed by the reaction. 

A Monte Carlo method was employed to investigate the uncertainties arising from optical model parameters. However, that method requires some initial optical model parameters. The priors of $^3$He and deuteron potentials were calculated from the global optical model parameters presented in Liang \textit{et al.}~\cite{Liang09} and Daehnick \textit{et al.}~\cite{Daehnick80}. The bound state parameters were chosen to be consistent with those of Ref.~\cite{Marshall23}. For the zero-range coupling constant $D_0$, a constant value of $-172.8$ MeV$\cdot$fm$^{3/2}$\cite{PhysRev.149.791} was applied in this work. Table \ref{tab:params} shows the optical model parameters following optimization. The underlined global parameters in the table were found to have large impacts on our measured $C^2S$ and were constrained by our experimental data. %Note that the $^3$He imaginary volume potential was not included in our model.
% The underlined parameters in the table were adjusted to fit on our data using $\chi^2$-minimization. 

Theoretical cross sections for both the elastic scattering and the transfer channel could be calculated by FRESCO given one set of parameters. The relationship between the experimentally determined cross section and that calculated by FRESCO are: 
\begin{align}
&\left(\frac{d\sigma}{d\Omega}\right)_{\text{ES,exp}}=\eta\times\left(\frac{d\sigma}{d\Omega}\right)_{\text{ES,DWBA}}\notag\\
&\left(\frac{d\sigma}{d\Omega}\right)_{\text{TR,exp}}=\eta\times  C^2S_{d+p} \times C^2S_{\text{Na}+p} \times \left(\frac{d\sigma}{d\Omega}\right)_{\text{TR,DWBA}}\;
\label{eq:2}.
\end{align}
Where $\frac{d\sigma}{d\Omega}$ is the differential cross section. The subscript ES denotes the elastic scattering channel and TR denotes the transfer channel. `exp' refers to the experimental cross sections and DWBA is the calculated cross sections from FRESCO. $\eta$ is a common normalization factor, which depends on the solid angle of the spectrograph, and target composition. It was introduced because both the elastic scattering and transfer reaction cross sections are normalized to the elastic scattering cross sections measured by the monitor detector in the target chamber. $C^2S$ is the spectroscopic factor. $C^2S_{d+p}$ denotes the projectile system and is approximated to be $3/2$~\cite{C2Sd+p}. $C^2S_{\text{Na}+p}$ denotes the target system and is referred to as $C^2S$ throughout this paper . $C^2$ is the isospin Clebsch-Gordan coefficient, and $C^2$=1/2 for all the states in this analysis since they have isospins $T=0$ or $T=1$. 

\subsection{Markov Chain Monte Carlo DWBA}\label{subsec:BayesianMCMC}

In Ref.~\cite{Garrett78} and earlier studies, $C^2S$ values were extracted with a $\chi^2$ minimization to fit the data with optimized optical model parameters. However, the corresponding statistical uncertainties from such process are rarely carefully evaluated. In this work, all the statistical uncertainties to our reported $C^2S$ were included using Bayesian statistics and MCMC method. More details about the methodology can be found in Refs.~\cite{Marshall20} and \cite{Marshall23}.

Generally speaking, Bayesian statistics uses the data, $\boldsymbol D$, and Bayes’ theorem to update the uncertain model parameters (the `priors') to provide uncertainties on those parameters (i.e., the `posterior' distributions). According to Bayes’ theorem:
\begin{align}
&\mathbf{P(\theta|D)=\frac{P(D|\theta)P(\theta)}{P(D)}}
\label{eq:Bayes}.
\end{align}
Where $\mathbf{P(\theta|D)}$ is the posterior, interpreted as our updated knowledge of the model parameters $\boldsymbol \theta$ with our observed data $\boldsymbol D$. $\mathbf{P(D|\theta)}$ is the likelihood function, interpreted as the conditional probability distribution of the data with given parameters. This is often given by the $\chi^2$ between a model prediction and the data. $\mathbf{P(\theta)}$ is the prior probability of the model parameters. In practice, this represents probability density distributions of every uncertain parameter before any data are taken. $\mathbf{P(D)}$ is the evidence, interpreted as the probability distribution of the data \cite{van2021bayesian}. This final parameter is extremely costly to compute, but constitutes a normalization parameter that is calculated along with a MCMC process in practice.

\begin{table*}[t]
\caption{\label{tab:params}Initial optical model potential parameters. The underlined parameters were constrained by $\chi^2$ fits and Bayesian MCMC processes. $V_r$, $r_r$, $a_r$ and $W_i$, $r_i$, $a_i$ are the real and imaginary volume potential parameters. $W_D$, $r_D$ and $a_D$ are for imaginary surface potential. $V_{so}$, $r_{so}$ and $a_{so}$ are spin-orbit parameters. $r_c$ is the Coulomb radius, which was found to be insignificant after normalizing our elastic scattering data to Rutherford cross sections calculated using the same radius. Note that the $^3$He potential did not include imaginary volume terms.}
\begin{ruledtabular}
\begin{tabular}{cccccccccccc}
 Interaction&$V_r$&$r_r$&$a_r$&$W_i$&$W_D=\frac14W'\footnote{$W'$ was used instead of $W_D$ in an old formalism. They are commutable simply by a factor of 4.}$&$r_i=r_D$&$a_i=a_D$&$V_{so}$&$r_{so}$&$a_{so}$&$r_c$\\
 \hline
$^{23}$Na+$^3$He\footnote{Based on Ref.~\cite{Liang09}}&\underline{117.4}&\underline{1.18}&\underline{0.75}&&\underline{19.9}&\underline{1.20}&\underline{0.81}&2.08&0.74&0.78&1.289 \\
 $^{24}$Mg+$d$\footnote{Based on Ref.~\cite{Daehnick80}}&\underline{87.6}\footnote{These initial parameters are dependent on the energy $E_x$ of the $^{24}$Mg excited state. Such dependency is insignificant compared with the constraint from MCMC. Listed values are calculated with $E_x=9516$ keV. }&\underline{1.17}&\underline{0.72}\footnotemark[4]&\underline{0.38}\footnotemark[4]&\underline{12.3}\footnotemark[4]&\underline{1.32}&\underline{0.73}\footnotemark[4]&6.88&1.07&0.66&1.30 \\
 $^{23}$Na+$p$& \footnote{Adjusted to match separation energy}&1.25&0.65&&&&&6.24&1.25&0.65&1.25 \\
\end{tabular}   
\end{ruledtabular}
\end{table*}

% The fitting was performed using the pfunk MCMC package \cite{pfunk} and follows the procedure described in Ref.~\cite{Marshall20}. Two major modifications were made in this work compared with Ref.~\cite{Marshall20}: First, a mixing ratio $\alpha$ is added with a uniform prior between (0,1) for the states with mixed angular momentum. Second, a percentage uncertainty $f_{elastic}$ is added to the elastic scattering cross section under the same consideration for the transfer channel.
To obtain the probability distribution for the spectroscopic factor, $C^2S$, first, all the uncertain parameters, including optical model parameters, were defined as informative prior probability distributions. Normal distributions centered at the calculated initial values were defined for some influential optical model parameters for $^3$He and deuteron potentials. Those parameters are underlined in Table.\ref{tab:params}. The standard deviations were initialized as 20\% and 10\% for $^3$He and deuteron, respectively, which represent the observed spread in the global optical models in this mass region. A half normal distribution with $\sigma=1.5$ was used for the $C^2S$ prior. The value $\sigma=1.5$ is a reasonable upper limit for experimentally measured $C^2S$ values. A normally-distributed parameter $D_0$ was added with $\sigma=15\%$ to account for the observed difference from different models according to Refs \cite{osti_4462813,Bertone2002}. Moreover, due to the discrete V-r ambiguity in the $^3$He model \cite{Marshall23}, a constraint $c$ is added so that $V_rr_r^{1.14}=c$ holds. $c$ is uniformly distributed centering at $c_0=132.9$ in a $\pm30\%$ interval.

\begin{figure}[ht]
\includegraphics[width=80mm]{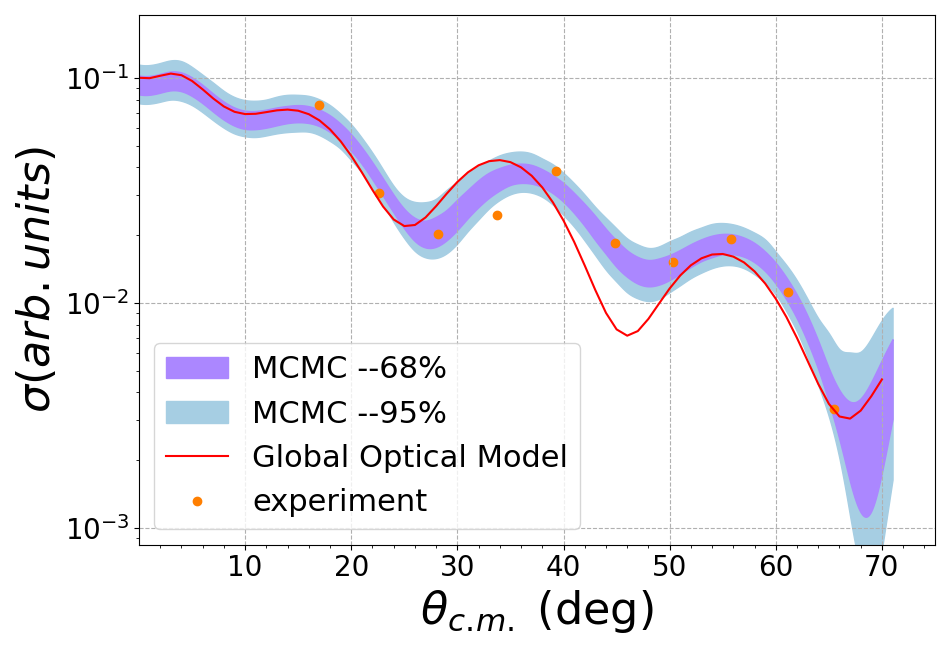}% Here is how to import EPS art
\caption{\label{fig:ES_MCMC} Experimental $^{23}$Na$(^3$He,$^3$He$)$$^{23}$Na cross sections normalized to Rutherford scattering cross sections, which is deduced from the yield of the monitor detector. }
\end{figure}

\begin{figure}[ht]
     \begin{subfigure}[l]{0.5\textwidth}
         \centering
         \includegraphics[width=80mm]{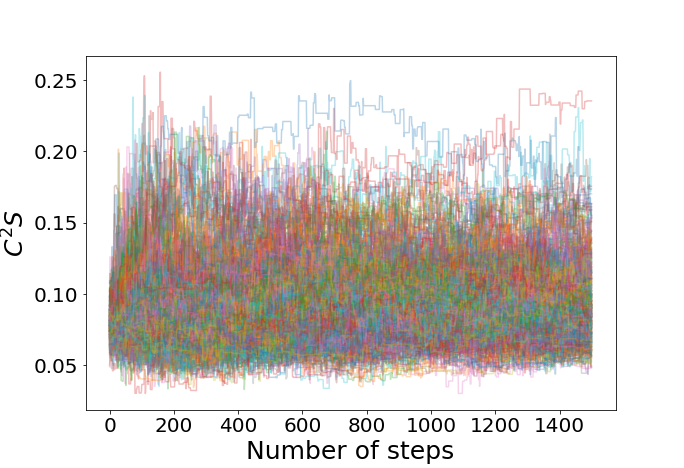}
     \end{subfigure}
     
     \begin{subfigure}[l]{0.5\textwidth}
         \centering
         \includegraphics[width=72mm]{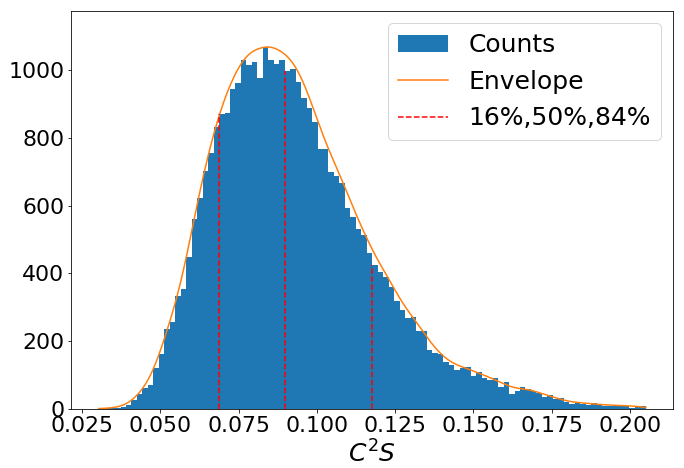}
     \end{subfigure}
\caption{\label{fig:C2S_MCMC} The Markov chain(top) and posterior(bottom) for the $C^2S$ of level 9516 keV. The 16\%, 50\% and 84\% of the posterior corresponding to the 68\% credible interval are also shown.}
\end{figure}

We used the nested sampling algorithm in the python package \textit{dynesty} \cite{dynesty} to draw samples and the python package \textit{emcee} \cite{emcee} for MCMC ensemble sampling. For each state, an ensemble of 400 random walkers was drawn from the prior distributions and run in 8000 steps to obtain the posteriors for the parameters. In each step the likelihood is evaluated based on the difference between the measured data and updated calculation from FRESCO for both channels. The first 6000 steps were discarded to ensure stability in the solutions. The last 2000 samples were then thinned by 50, which gives 16,000 samples with 400 walkers in total. Such sampling process takes $\sim$3 days for each state. Fig. \ref{fig:ES_MCMC}. shows the MCMC results for our elastic scattering data. The 95\% and 68\% credible intervals from the MCMC are shown in light and dark blue. The calculated cross sections with the global parameters from Table. \ref{tab:params} is also shown. %The $^3$He potential parameters were found to be significant mostly for the elastic scattering channel. A similar plot as Fig. \ref{fig:ES_MCMC} was obtained for each state but not presented in this paper. 
For the transfer reaction data analysis, a sample of Markov chain and a log-normal shaped posterior for the $C^2S$ are shown in Fig. \ref{fig:C2S_MCMC}. The 16\%, 50\% and 84\% percentiles of the posterior corresponding to the 68\% credible interval are also shown. 

% \begin{eqnarray}
% \sigma'^2_{elastic,i}=\sigma^2_{elastic,i}+\left(f_{elastic}\frac{d\sigma'}{d\Omega}_{optical,i}\right)
% \label{eq:one}.
% \end{eqnarray}

\subsection{Resolving Doublets with MCMC}\label{subsec:ResolveMCMC}

A specific challenge in this analysis arises from the state near 10.7 MeV, which appears as a doublet with a reported energy separation of 19 keV, comparable to the $\sim$20 keV resolution(FWHM) of our focal-plane detector. Due to this, conventional fitting could not reliably resolve the two components. Given the significance of this doublet, which contributes approximately 23\% to the total direct capture reaction rate, prior information on peak shapes and widths was used to extract their individual $C^2S$.

From the known energies and uncertainties of this doublet and the energy calibration of the focal-plane detector described above, a Bayesian model was constructed to simultaneously extract the areas of each peak in the doublet. In this model, two Gaussian peaks were defined with informative priors on their positions, as constrained by the energy and calibration uncertainties. The peaks were defined with a common width, again constrained by nearby peaks in the focal-plane spectrum. The background shape was described as a linear model as used for all other peaks in the present analysis. This method yielded peak areas for both peaks in the doublet. It was tested for a range of conditions and found to be accurate within the reported uncertainties.

\section{Results and Discussion}

\subsection{Spectroscopic Factors}\label{subsec:results}
The differential cross sections and MCMC results for $^{24}$Mg states are presented in Fig. \ref{fig:MCMC results}.  
 The spectroscopic factors obtained for each of those states are summarized in Table. \ref{tab:summary of C2S}. A column of single-particle asymptotic normalization coefficients (ANCs) is also presented for completeness. More details about the ANCs are explained in detail in Sec. \ref{subsec:ANC_exp}. The listed uncertainties of the $C^2S$ correspond to 68\% confidence interval from the MCMC, which include uncertainties from: (i) variation of the optical-model parameters; (ii) the strength parameter $D_0$; (iii) measured yields, which typically have small uncertainties as long as there were no contaminant peaks; and (iv) ambiguities in the assignments of the transfer orbital angular momentum. In addition, a systematic uncertainty of 18\% should be considered for the final $C^2S$ values in this work, to account for a difference for normalizing the monitor detector yield between elastic scattering and transfer measurements. Tentatively assigned spin, parity and orbital configuration are listed in parentheses in Table \ref{tab:summary of C2S}. For each level, when multiple orbital configurations are possible, the recommended configuration is listed on the first line in regular font, while alternative configurations and their corresponding $C^2S$ values are shown on the second line in italics.
\begin{table*}[htbp]
\caption{Summary of spectroscopic factors.}
\begin{center}
\renewcommand{\arraystretch}{1.3}
\begin{tabular}{p{0.1\textwidth}p{0.08\textwidth}p{0.15\textwidth}p{0.23\textwidth}p{0.13\textwidth}p{0.1\textwidth}}
  \hline
  \hline
  $E_x(\text{keV})$& $J_f^\pi$ \footnotemark[1] & Orbital &  \multicolumn{2}{c}{$(2J_f+1)C^2S$}  & ANC($C_{24}$) \\
    && &This work&Literature  & (fm$^{-1/2}$)\\
  \hline
$7349$ & $2^+$ & $2s_{1/2}$+$1d_{5/2}$ & $0.024^{+0.010}_{-0.006}+0.011^{+0.009}_{-0.007}$& 0.09+0.10\footnotemark[2] & -1.065, 0.28\\ 
&  & \textit{2s}$_{1/2}$ & $\mathit{0.025}$\footnotemark[3] & 0.04\footnotemark[4] & \textbf{--}\\ 
$7555$ & $1^-$ & $1p_{1/2}$ & $0.009^{+0.005}_{-0.003}$ & \textbf{--} & 0.42\\ 
 & & \textit{2p}$_{3/2}$ & $\mathit{0.0045}\footnotemark[3]$ & 0.02\footnotemark[2] & -0.50\\
$7748$ & $1^+$ & $2s_{1/2}$+$1d_{5/2}$ & $0.097^{+0.038}_{-0.026}+0.137^{+0.067}_{-0.045}$ & 0.32+0.58\footnotemark[2] & -2.90, 0.94\\ 
 &   & \textit{2s}$_{1/2}$ &  $\mathit{0.162}$\footnotemark[3]  & 0.11\footnotemark[4] & \textbf{--}\\
$8358$ & $3^-$ & $1p_{3/2}$ & $0.061^{+0.019}_{-0.013}$ & \textbf{--} & 0.70\\
 &  & \textit{2p}$_{3/2}$ & $\mathit{0.063^{+0.020}_{-0.014}}$ & 0.08\footnotemark[2]&  -1.07\\[0.3em]
% \{\makecell[l]{8.438 \\8.439 } &\makecell[l]{$1^-$ \\$4^+$ } & 2p3/2+$1d_{5/2}$ & $0.21^{+0.12}_{-0.09}+0.53^{+0.25}_{-0.17}$ & 0.28+1.04\footnotemark[2] & 0.56, -2.55\\
%   & & 1p1/2+$1d_{5/2}$ & $0.063^{+0.031}_{-0.024}+0.44^{+0.19}_{-0.12}$ & \textbf{--} & 1.80, 0.93\\ 
\{\makecell[l]{8438 \\[0.2em]8439 } &\makecell[l]{$1^-$ \\[0.2em]$4^+$ } &\makecell[l]{$2p_{3/2}+1d_{5/2}$\\[0.2em]\textit{1p}$_{1/2}$+\textit{1d}$_{5/2}$} & \makecell[l]{$0.071^{+0.037}_{-0.026}+0.51^{+0.23}_{-0.15}$ \\[0.2em]$\mathit{0.19^{+0.14}_{-0.09}+0.50^{+0.23}_{-0.15}}$}&  \makecell[l]{0.28+1.04\footnotemark[2]\\[0.2em]\textbf{--}} & \makecell[l]{-1.77,0.92\\[0.2em] -2.91,0.91}\\[1em]
$8655$ & $2^+$ & $2s_{1/2}$ + $1d_{5/2}$ & $0.117^{+0.050}_{-0.029}+0.114^{+0.061}_{-0.042}$ & 0.34+0.41\footnotemark[2] &-2.21, 0.56\\ 
  &   & \textit{2s}$_{1/2}$  & $\mathit{0.18}$\footnotemark[3] & 0.11\footnotemark[4] &\textbf{--} \\ 
$8864$ & $2^-$ & $1p_{1/2}$ & $0.33^{+0.13}_{-0.09}$ & 0.39\footnotemark[2]& 1.67\\
 &  & \textit{2p}$_{3/2}$ & $\mathit{0.13^{+0.04}_{-0.03}}$ & 0.135\footnotemark[4]& -1.81\\
$9003$ & $2^+$ & $2s_{1/2}$ & $0.016^{+0.008}_{-0.005}$ &\textbf{--}& -14.05\\
 &  & \textit{1d}$_{5/2}$ & $\mathit{0.026^{+0.011}_{-0.008}}$ &\textbf{--}& 0.25\\
$9146$ & $1^-$ & $1p_{1/2}$ & $0.034^{+0.011}_{-0.008}$ &\textbf{--}& -0.71 \\
 &  & \textit{2p}$_{3/2}$ & $\mathit{0.014^{+0.005}_{-0.003}}$ &\textbf{--}& 0.67 \\
\{\makecell[l]{9284 \\[0.2em]9301 } & \makecell[l]{$2^+,4^+$  \\[0.2em]$(4^+)$  } & $1d_{5/2}$ & $0.258^{+0.081}_{-0.058}$ & 0.58\footnotemark[2] &0.55\\[0.4em]
$9458$ & $(3)^+$ & $1d_{5/2}$ & $0.220^{+0.073}_{-0.050}$ & 0.49\footnotemark[2] & 0.56\\ 
$9516$ & $4^+$ & $1d_{5/2}$ & $0.48^{+0.14}_{-0.10}$ & 1.2\footnotemark[2], 0.585\footnotemark[4]& 0.73\\ 
$9828$ & $1^+$ & $1d_{5/2}$ & $0.245^{+0.077}_{-0.055}$ &\textbf{--} &0.84\\ 
 &   & \textit{2s}$_{1/2}$+\textit{1d}$_{5/2}$ & $\mathit{0.010^{+0.010}_{-0.007}+0.202^{+0.053}_{-0.039}}$  & (0.04)+0.54\footnotemark[2]&-0.78, 0.76\\[0.2em]
$9965$ & $1^+$ & ($2s_{1/2}$+$1d_{5/2}$) & $0.110^{+0.120}_{-0.046}+0.365^{+0.755}_{-0.142}$ & 0.18+0.31\footnotemark[2]&-2.60, 1.00\\ 
$10028$ & $5^-$ & ($1f_{7/2}$) & $0.223^{+0.083}_{-0.060}$ &\textbf{--}&0.12\\ 
$10059$ & $(1,2)^+$ & $2s_{1/2}$+$1d_{5/2}$ & $0.126^{+0.062}_{-0.039}+0.41^{+0.20}_{-0.13}$ & \makecell[l]{0.22+0.72\footnotemark[2], \\0.08+0.25\footnotemark[4]  }& -2.89, 1.27\\
$10161$ & $(0^+)$ & ($1d_{3/2}$) & $0.31^{+0.10}_{-0.07}$ & \textbf{--} &1.45\\ 
$10334$ & $(3^-)$ & ($1p_{3/2}$) & $0.013^{+0.006}_{-0.004}$ &\textbf{--}& 0.29 \\
$10361$ & $2^+$ & $1d_{5/2}$ & $0.266^{+0.079}_{-0.058}$ & 1.12\footnotemark[2] & 0.63\\ 
$10660$ & $(3^+,4^+)$ & ($1g_{7/2}$) & $0.159^{+0.053}_{-0.039}$ & \textbf{--} & 0.02\\ 
$10712$ & $1^+$ & $2s_{1/2}$ & $0.33^{+0.09}_{-0.12}$ & 0.53\footnotemark[5] &-4.51\\ 
&  & \textit{2s}$_{1/2}$+\textit{1d}$_{5/2}$ & $\mathit{0.16^{+0.11}_{-0.07}+0.08^{+0.06}_{-0.05}}$ & 0.38+0.16\footnotemark[5]&-4.17, 0.38\\ 
$10731$ & $2^+$ & $2s_{1/2}$ & $0.37^{+0.10}_{-0.08}$ & 0.84\footnotemark[5] &-4.34\\ 
&  & \textit{2s}$_{1/2}$+\textit{1d}$_{5/2}$ & $\mathit{0.35^{+0.21}_{-0.17}+0.09^{+0.11}_{-0.10}}$ & 0.72+0.12\footnotemark[5]&-3.67, 0.46\\
$10821$ & $3^+,4^+$ & $1d_{5/2}$ & $0.027^{+0.008}_{-0.006}$ & 0.13\footnotemark[2], 0.057\footnotemark[5] &0.17\\ 
& & \textit{1d}$_{5/2}$+\textit{1g}$_{7/2}$ & $\mathit{0.030^{+0.018}_{-0.008}+0.004^{+0.030}_{-0.013}}$ & 0.05+0.035\footnotemark[5] &0.19, 0.002\\ 

$10917$ & $2^+$ & $1d_{5/2}$ & $0.079^{+0.025}_{-0.019}$ & 0.18\footnotemark[2], 0.14\footnotemark[5] & 0.37\\ 
$11217$ & $3^+,4^+$ & $1d_{5/2}$ & $0.84^{+0.26}_{-0.19}$ & 0.8\footnotemark[6] & 1.21 \\
$11314$ & $(3,4)^+$ & $1d_{5/2}$ & $0.30^{+0.09}_{-0.06}$ & 0.48\footnotemark[5] & 1.04\\
$11391$ & $1^-$ & $1p_{1/2}$ & $0.12^{+0.04}_{-0.03}$ & \textbf{--} & 4.28\\ 
&  & \textit{2p}$_{3/2}$ & $\mathit{0.054^{+0.018}_{-0.012}}$ & 0.066\footnotemark[7] , 0.06\footnotemark[5] &-4.34\\
\hline
\hline
\end{tabular}
\label{tab:summary of C2S}
\end{center}
\end{table*}

\addtocounter{table}{-1}

\begin{table*}[htbp]
\caption{Summary of spectroscopic factors (Continued).}
\begin{center}
\renewcommand{\arraystretch}{1.3}
\begin{tabular}{p{0.1\textwidth}p{0.08\textwidth}p{0.15\textwidth}p{0.23\textwidth}p{0.13\textwidth}p{0.1\textwidth}}
  \hline
  \hline
  $E_x(\text{keV})$& $J_f^\pi$ \footnotemark[1] & Orbital &  \multicolumn{2}{c}{$(2J_f+1)C^2S$}  & ANC($C_{24}$) \\
    && &This work&Literature  & (fm$^{-1/2}$)\\
  \hline
\{\makecell[l]{$11453$ \\ $11457$ } & \makecell[l]{$2^+$ \\$(0^+)$ } & $2s_{1/2}$+$1d_{5/2}$ & $0.128^{+0.064}_{-0.039}+0.070^{+0.041}_{-0.028}$ & \makecell[l]{0.14+0.05\footnotemark[7]\\0.24+0.16\footnotemark[5]} &-14.74, 1.32\\[0.8em]
\{\makecell[l]{$11522$ \\ $11527$ } & \makecell[l]{$2^+$\\$(2^+)$} & ($2s_{1/2}$+$1d_{5/2}$) & $0.095^{+0.167}_{-0.030}+0.046^{+0.033}_{-0.022}$ & \makecell[l]{0.05+0.057\footnotemark[7] \\ (0.10)\footnotemark[5] }&-26.48, 2.20\\[0.5em]
$11698$ & $4^+$ & $1d_{5/2}$ & $0.076^{+0.024}_{-0.017}$ & 0.085\footnotemark[7], 0.11\footnotemark[5]& \textbf{--}\\ 
$11933$ & $(3)^+$ & $1d_{5/2}$ & $0.19^{+0.056}_{-0.041}$ & 0.23\footnotemark[7], 0.25\footnotemark[5]& \textbf{--}\\ 
\hline\hline
\end{tabular}
\label{tab:summary of C2S(continued)}
\footnotetext[1]{From ENSDF \cite{ENSDFNOV22}}
\footnotetext[2]{From Garrett \textit{et al.} Ref.~\cite{Garrett78}.}
\footnotetext[3]{\parbox [t] {2\linewidth} {\hspace{-3.1cm}From simple $\chi^2$ fit just for comparing with old literature. The uncertainty is estimated to be 33\%. }}
\footnotetext[4]{From Tang \textit{et al.} Ref.~\cite{Tang}.}
\footnotetext[5]{From Hale \textit{et al.}~\cite{Hale04}}
\footnotetext[6]{From Fuchs \textit{et al.}~\cite{Fuchs}}
\footnotetext[7]{%
  \makebox[2\textwidth][l]{%
    \begin{minipage}[l]{0.9\textwidth}
    \raggedright
      From Marshall \textit{et al.} Ref.~\cite{Marshall23}. For those states, our $(2J+1)C^2S$ values represent a re-analysis of Marshall's to verify consistency and should not be considered new measurements.
    \end{minipage}%
  }%
}
\end{center}
\end{table*}

% It was found that for two angular momentum configurations (whether in one mixed state or in two separate states) with spectroscopic factors ($C^2S$) of the same order of magnitude, an increase in the $l$ value by 2 is associated with a reduction of the experimental S-factor by approximately one order of magnitude. 

When comparing the results presented here with other experimentally-determined spectroscopic factors, it is important to take into account the effect of the bound state potential parameters. The parameters used in this work are $r_r$=1.25 fm and $a_r$=0.65 fm following Ref.~\cite{Marshall23}. To investigate the effect of the bound-state model parameters, the 9516 keV was considered. Varying the radius $r_r$ by 10\% results in a $\sim$40\% change in $C^2S$ and $\sim$20\% for the single-particle ANC. Varying $a_r$ by 10\% changes $C^2S$ and ANC by $\sim$14\% and $\sim$7\%, respectively. Note that the direct capture calculation must be performed with consistent bound state parameters, which largely cancels out these uncertainties as discussed in Refs.~\cite{Bertone2002}, \cite{Marshall20}, and \cite{Harrouz2022}.

\begin{figure*}[ht]
    \centering
      \begin{subfigure}[b]{0.47\textwidth}
         \centering
         \includegraphics[width=\textwidth]{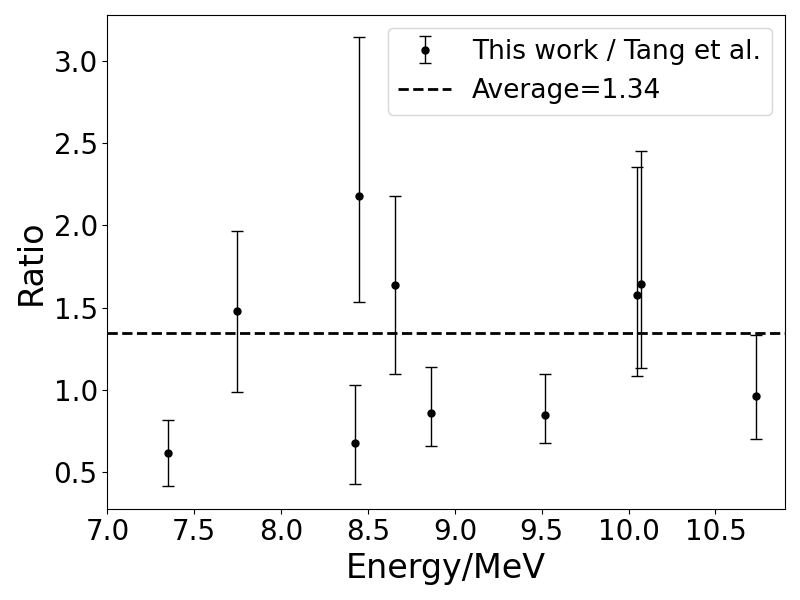}
     \end{subfigure}
     % \hfill
     \begin{subfigure}[b]{0.47\textwidth}
         \centering
         \includegraphics[width=\textwidth]{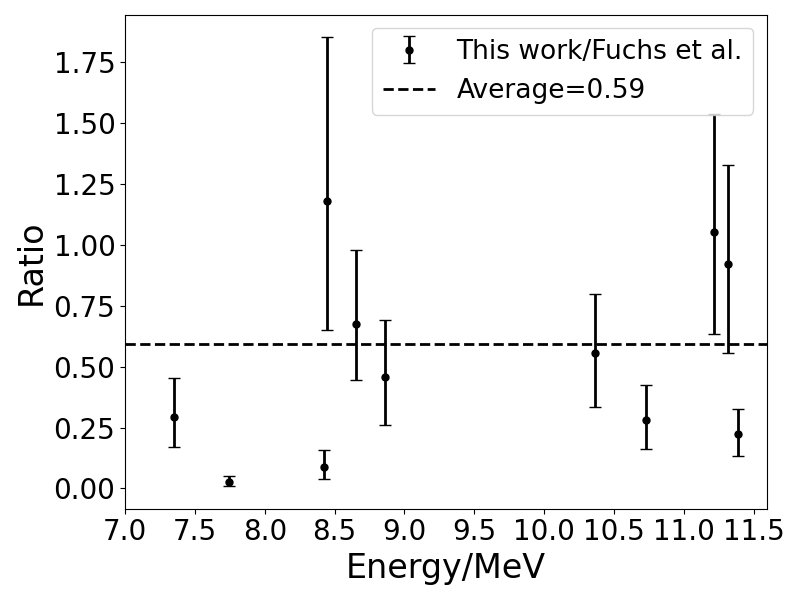}
     \end{subfigure}
     \\
     \begin{subfigure}[b]{0.47\textwidth}
         \centering
         \includegraphics[width=\textwidth]{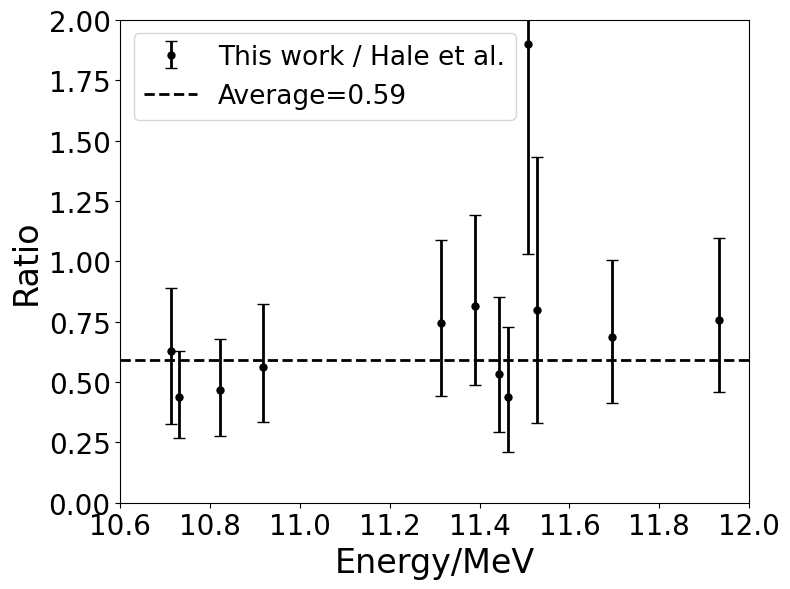}
     \end{subfigure}
     % \hfill
     \begin{subfigure}[b]{0.47\textwidth}
         \centering
         \includegraphics[width=\textwidth]{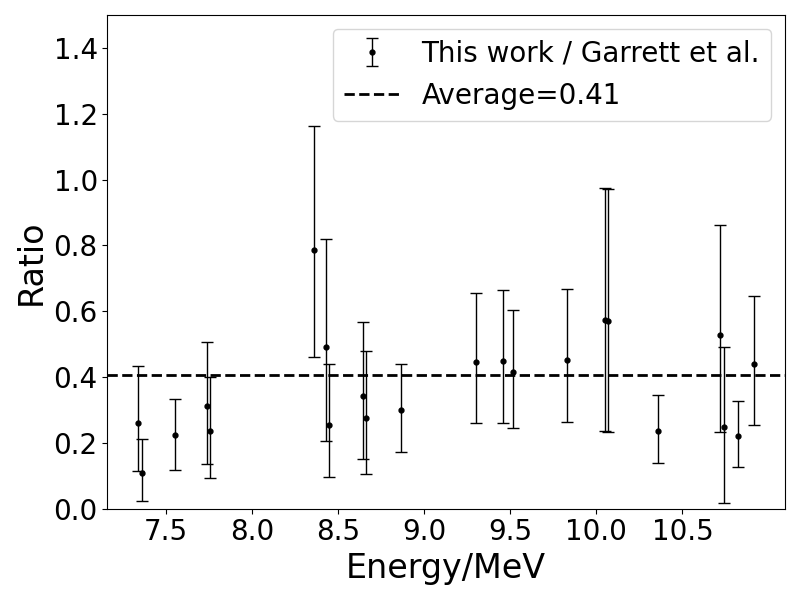}
     \end{subfigure}
\caption{\label{fig:C2S_Ratio} The ratio of $(2J_f+1)C^2S$ values between this work and literature, showing the degree of agreement. Mixed angular momentum components are separated for 20 keV in the plot for visualization. }
 % Left: Garrett et al. \cite{Garrett78}. Right: Hale et al. \cite{Hale04}. 
\end{figure*}

A comparison between $(2J_f+1)C^2S$ derived from this work and literature is shown in Fig. \ref{fig:C2S_Ratio}, plotted as ratios of (this work/literature). Note that a systematic difference of up to $\sim50\%$ is expected when comparing $C^2S$ values from $(^3\text{He},d)$ and $(d,n)$ studies, due to the different optical model parameters and reaction mechanisms involved. Tang \cite{Tang}, Hale \cite{Hale04} and Garrett \cite{Garrett78} used \texttt{DWUCK4} for their DWBA calculation. Fuchs \cite{Fuchs} used an older DWBA code \texttt{SALLY}. The values used for plotting are given in Table \ref{tab:summary of C2S}. A simple average of the ratio among the levels is plotted as dashed line to show the systematic differences between datasets. We found that our $C^2S$ values are mostly consistent with Tang \textit{et al.}~\cite{Tang} and Hale \textit{et al.}~\cite{Hale04}, but are smaller by $\sim$60\% on average compared with Fuchs \textit{et al.}~\cite{Fuchs} and Garrett \textit{et al.}~\cite{Garrett78}. The latter study has been used exclusively for calculating the $^{23}$Na(p,$\gamma$)$^{24}$Mg direct capture cross sections in previous studies. We performed a number of additional calculations and found that the different choice of optical model parameters between Ref.~\cite{Garrett78} and our work only explains less than 20\% of the difference. The remaining systematic difference likely stems from systematic experimental effects. Since we use spectroscopic factors from Garrett for states below our experimental region, those spectroscopic factors are scaled by a factor of 0.4 to ensure consistency with our measured states in the direct capture reaction rate calculation.

% More details are discussed in the next subsection (\ref{subsec:compare}).

% We believe that the observed difference is mainly caused by their absolute cross section determination(they assigned 30\% systematic uncertainty mainly from their target thickness). some difference brought by their experimental techniques(using nuclear emulsion) or bound state parameters. The bound state potential parameters used in these studies are listed in Table \ref{tab:boundparams} together with the overall systematic differences of their $C^2S$ from this work. We give an estimation of $<10\%$ systematic difference on the $C^2S$ from bound state parameters for Garrett \textit{et al.}

% \begin{table}[t]
% \caption{\label{tab:boundparams}Bound state parameters from references for DWBA analysis}
% \begin{ruledtabular}
% \begin{tabular}{cccc}
%  Ref.&$r_r$&$a_r$&Differences in $C^2S$\\
%  \hline
% This work. &1.25&0.65&-\\
% Tang \textit{et al.} &1.25&0.65&+26\%\\
% Hale \textit{et al.}&1.17&0.69&-27\%\\
% Garrett \textit{et al.}&1.26&0.60&-59\%\\
% \end{tabular}   
% \end{ruledtabular}
% \end{table}
More details about our measured states are given below. The angular momentum assignments are consistent with Garrett \textit{et al.}~\cite{Garrett78} unless otherwise noted. 

\textit{7349 keV, $2^+$}: This state was observed by both Garrett\cite{Garrett78} and Tang\cite{Tang}, with Garrett assigning a mixed $l = 0+2$ component and Tang suggesting a pure $\ell=0$ transfer. Our data support Tang. The S-factor contribution from $\ell=2$ is estimated to be less than 5$\%$ of that of the $\ell=0$ component from our measurements. This state is used as a calibration point due to its position at the low energy end of the spectra. 

\textit{7555 keV, $1^-$}: This state suffers from poor statistics in our data, with a signal-to-background ratio of approximately 1:1 across all angles. Consequently, data at 19° and 21° could not be extracted.

\textit{7748 keV, $1^+$}: This state was reported to be $\ell=0+2$ mixed transitions by Garrett \cite{Garrett78} and pure $\ell=0$ transitions by Tang \cite{Tang}. Our angular data support a mixed $\ell=0+2$ transfer, in line with Garrett.

\textit{8358 keV, $3^-$}: No output was obtained from FRESCO for the 1p${1/2}$ configuration. The 1p${3/2}$ orbital is allowed and yields a $C^2S$ value comparable to that of a $2p_{3/2}$ assignment.

\textit{8439 keV, $1^-$} and \textit{8439 keV, $4^+$}: These two states appear as an unresolved doublet. Angular data were fitted using $\ell=1+2$. Both $2p_{3/2}+1d_{5/2}$ and $1p_{1/2}+1d_{5/2}$ orbital combinations were considered. The former is used for comparison with Ref.~\cite{Garrett78}. 

\textit{8655 keV, $2^+$}: Also observed by Garrett and Tang, with the same disagreement in $l$ assignment as the 7748keV state. Our data favor an $\ell=0+2$ transfer.

\textit{8864 keV, $2^-$}: All observations and fits are in agreement with Garrett and Tang.

\textit{9003 keV, $2^+$}: This state was previously observed but weakly populated in Garrett’s work. In our experiment, the data show a likely contamination peak from an unknown source in the 11–15° angular range. An $\ell=0$ assignment yields the best fit, though results from an $\ell=2$ fit are also provided for comparison, assuming the 11–15° region is uncontaminated. The plot for this level in Fig.~\ref{fig:MCMC results} shows only the results within the $68\%$ credible interval for clarity. This work gives the first extraction of the $C^2S$ for this state. Despite the contamination, the extracted $C^2S$ is sufficiently small that this state contributes negligibly to the direct capture rate.

\textit{9146 keV, $1^-$}: Previously observed by Garrett with weak population. An unknown contaminant peak was present at 5–7°, but the data still show a clear $\ell=1$ feature. No previous measurements report $C^2S$ from this state.

\textit{9284 keV, $2^+$,$4^+$} and \textit{9301 keV, $(4^+$)}: This energy region includes three previously reported levels: a 9284 keV state assigned $J^\pi = 2^+, 4^+$, a 9301 keV state with $J^\pi = (4^+)$, and a 9299.8 keV state with unknown $J^\pi$. In our spectra, this appears as an unresolved multiplet with at least two components. The dominant contribution arises from an $\ell = 2$ transition, which is most consistent with the previously reported 9301 keV $(4)^+$ state. Accordingly, a $J^\pi = 4^+$ assignment was adopted for calculations in this work. A small $\ell = 0$ component may also be present, corresponding to a $J^\pi = (1$–$2)^+$ state at a lower excitation energy. A two-component fit in this scenario yields $C^2S_{(\ell = 0+2)} = 0.0066 + 0.252$. However, considering the range of uncertainties in our data, the presence of such a component is not required and may not be physically significant, as the fit remains dominated by the $\ell = 2$ contribution.

\textit{9458 keV, $(3)^+$}: This state is affected by contamination from the $^{14}$N*(5106 keV) peak between 13–17°. There was a tentative spin assignment from Ref.\cite{Paper12C16Oag}, with positive parity supported by both prior $(^3\text{He},d)$ and $(^3\text{He},^3\text{He})$ data and confirmed in this work.

\textit{9516 keV, $4^+$}: Appears as a multiplet across all angles. The dominant component corresponds to a known $4^+$ state from $^{20}\text{Ne}(\alpha,\gamma)$\cite{FIFIELD197877}, with unresolved contributions from possible $(6^+)$ and $(2,3)^+$ states.

\textit{9828 keV, $1^+$}: Garrett assigned $\ell=0+2$ with a weak component, but our data favor a pure $\ell=0$ transfer. 

\textit{9965 keV, $1^+$, 10028 keV, $5^-$}, and \textit{10059 keV, $(1,2)^+$}: These states are affected by contamination from the $^{14}$N ground state and are difficult to resolve. The missing data in the middle of their angular range result from this contamination. Nevertheless, an $\ell = 0+2$ transfer best describes the 9965 keV and 10059 keV data. For the 10028 keV state, an $\ell = 3$ transfer and a 1f$_{7/2}$ orbital were tentatively assigned, consistent with its known $J^\pi = 5^-$. The 10028 keV level is observed here for the first time in the $(^3\text{He},d)$ reaction. The 10059 keV state was mostly resolved from the 10028 keV doublet. An $\ell = 0+2$ fit constrains its $J^\pi$ to $(1,2)^+$. A previous $^{20}$Ne$(\alpha,\gamma)$ study by Schmalbrock \textit{et al.}~\cite{SCHMALBROCK1983279} assigned a $2^+$, but their limited data points and large uncertainties in their $A_2$ values prevented a definitive assignment. Additional references for this state can be found in Table 24.22.a of Ref.~\cite{ENDT19901}. For the reaction rate calculations in this work, a $J^\pi = 1^+$ assignment was adopted.

\textit{10161 keV, $(0^+)$}: The data for this state was contaminated by the $^{14}$N*(5834 keV) peak. Garrett \cite{Garrett78} assigned an $\ell = 1$ transfer, although they noted that an $\ell = 2$ fit produced similarly good agreement. Our data support an $\ell = 2$ transfer, consistent with a possible $J^\pi = (0$–$4)^+$. The only existing $J^\pi$ information comes from a tentative $(0^+)$ assignment in a (p,p$'$) study~\cite{24Mgpp}, based solely on features of the angular distribution without quantitative fitting. A $J^\pi = 0^+$ assignment and a $1d_{3/2}$ orbital provides a good fit to our data. If the 1d$_{5/2}$ orbital is used—corresponding to $J^\pi = (1$–$4)^+$—the resulting $(2J+1)C^2S$ value is approximately 20% smaller.

\textit{10334 keV, $(3^-)$}: This state is weakly populated and appears at the tail of the 10361 keV peak. It is observed here for the first time in a transfer reaction. An $\ell = 1$ transfer provides a good fit to our data, confirming a negative parity consistent with previous $(\alpha,\gamma)$ studies. The 1p$_{3/2}$ orbital was adopted in our analysis. 

\textit{10361 keV, $2^+$}: All observations and fits are in agreement with Garrett \textit{et al.}~\cite{Garrett78}.

\textit{10660 keV, $(3^+,4^+)$}: Our data suffer from low counting statistics. $\ell=4$ has been reported in (p,p') \cite{24Mgpp} and fits well with our data.
Angular data were measured by Garrett, but not fitted. Our data also show very weak population with a 1:1 signal-to-background ratio. $\ell=4$ gives a good fit, but the assignment is tentative due to low statistics and possible interfering states nearby with $J^\pi=(1,2^+)$ and $J^\pi=0^+$. The $J^\pi=(3^+,4^+)$ assignment is favored based on evidence from $^{20}\text{Ne}(\alpha,\gamma)$\cite{FIFIELD197877}. $J^\pi=(3^+)$ was used for the rate calculations in this work.

\textit{10712 keV, $1^+$ and 10731 keV, $2^+$ states.} These two levels contribute significantly (approximately 23\%) to the total direct capture (DC) reaction rate (see Sect.~\ref{sec:DC}). They were resolved in Hale \textit{et al.}\cite{Hale04} but not in Garrett\textit{et al.}~\cite{Garrett78}. In the present experiment, their separation approached the resolution limit of the focal-plane detector. The doublet was resolved using Bayesian MCMC techniques (see Sect.\ref{subsec:ResolveMCMC} for details). Results from both $\ell = 0$ and $\ell = 0+2$ are presented for the resolved data. To compare with literature values based on the unresolved data, additional fit treating the peak as unresolved was carried out using $J^\pi=1^+$. The corresponding $(2J+1)C^2S$ was $0.37^{+0.10}_{-0.17}$ for a pure $2s_{1/2}$ orbital and $0.37^{+0.14}_{-0.09}+0.07^{+0.06}_{-0.06}$ for a mixed $2s_{1/2}+1d_{5/2}$ orbital. The corresponding ANC values are -5.53 for the 2$s_{1/2}$ component and 0.11 for the 1$d_{5/2}$ component.

\textit{10821 keV, $3^+,4^+$}: This state has been weakly populated and could not be resolved from the contaminant $^{14}N-$ 6446keV state at 13-17°. $\ell=2$ fitted well and is consistent with Garrett. Hale proposed an additional $\ell=4$ component to account for excess counts at large angles, which is not observed in this work. We confirmed the positive parity and adopted $J^\pi=3^+$ based on $K^\pi=1^+$ band predictions from Garrett.

\textit{10917 keV, $2^+$}: All observations and fits are in agreement with Garrett and Hale.

\textit{11217 keV $3^+,4^+$}: This state was contaminated by the $^{17}F$ ground state peak in 5–17° but was successfully resolved. $\ell=2$ describes our data well, suggesting $J^\pi=(0-4)^+$. Fuchs \textit{et al.}~\cite{Fuchs} reported 11.222keV 4+ state and a 11.223 $J^\pi=$(1,2)+ state, which is not accepted in ENSDF. Our calibrated energy 11202.4(34) keV(from the uncontaminated angular region), along with the observed doublet-like structure in the spectrum, indicates additional contributions at lower excitation energy. One candidate is the 11181 keV state reported in the $(p,p')$ study of Ref.\cite{24Mgpp}. Since the spins and parities are not clear for the lower lying contributions, a tentative assignment of $J^\pi=4^+$ was used for the calculations in this analysis.

%Data in the range of 5-13 degree were resolved by calculating and fixing the center of the Gaussian using calibrated energies from 15-21 degree. The provided error bars in Fig.\ref{fig:MCMC results} at 5-13 degree were supposed to be larger, but the actual uncertainties in those region were hard to estimate. 

\textit{11314 keV, $(3,4)^+$}: All observations and fits are in agreement with Hale. $J=3$ was used for the calculations.

\textit{11391 keV, $1^-$}: All observations and fits are consistent with Hale \textit{et al.}~\cite{Hale04} and Marshall \textit{et al.}~\cite{Marshall23}. As reported by Marshall, only the state with $J^\pi = 1^-$ is considered to be populated based on the $\ell = 1$ fit of our data..

\textit{11453 keV, $2^+$} and \textit{11457 keV, $(0^+)$}: Four near-threshold states at this energy and above were analyzed for comparison with results from Marshall\cite{Marshall23}, with good agreement in the extracted $C^2S$ values. This state was originally reported by Hale as an unresolved doublet consisting of $2^+$ and $(0^+)$ components. An $\ell = 0+2$ fit best describes our data. As discussed by Marshall\cite{Marshall23}, the $2^+$ state is likely the dominant component. Furthermore, an $\ell = 0$ transition would not be consistent with population of a $0^+$ state in this reaction.

\textit{11522 keV, $2^+$} and \textit{11527 keV, $(2^+)$}: Observed by Hale with both $\ell=0+2$ and $\ell=0+1+3$ fits. Our data support an $\ell=0+2$ assignment. The discrepancy with the $C^2S$ values of Marshall~\cite{Marshall23} is likely due to the inclusion of data points at higher angles in the present work.

\textit{11698 keV, $4^+$}: Consistent with Hale and Marshall. The peaks at 5–9° is contaminated by the first excited state of $^{17}F$ and could not be resolved. This state and the 11933 keV state were not included in our DC reaction rate calculations since they were above the proton threshold.

\textit{11933 keV, $(3)^+$}: All observations and fits are in agreement with Marshall \textit{et al.}~\cite{Marshall23}.

% MCMC figure
\begin{figure*}
     \centering
  \begin{minipage}[c]{0.035\textwidth}
    \centering
    \rotatebox{90}{\textbf{\normalsize $\sigma$ (arb.units)}}
      \end{minipage}%
        \hspace*{-0.4cm}  % bring subfigures closer
  \begin{minipage}[c]{0.96\textwidth}
    \centering
     \begin{subfigure}[l]{0.3\textwidth}
         \centering
         \includegraphics[ width=\textwidth,trim={40 50 0 0}, clip]{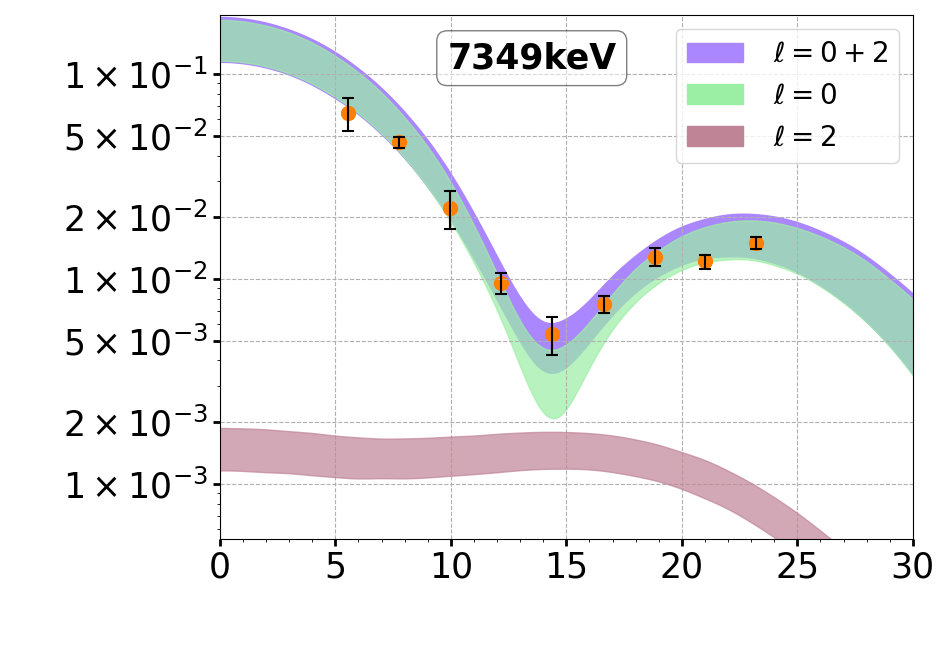}
         \label{fig:7349_twol}
     \end{subfigure}
     \begin{subfigure}[l]{0.3\textwidth}
         \centering 
         \includegraphics[ width=\textwidth,trim={40 50 0 0}, clip]{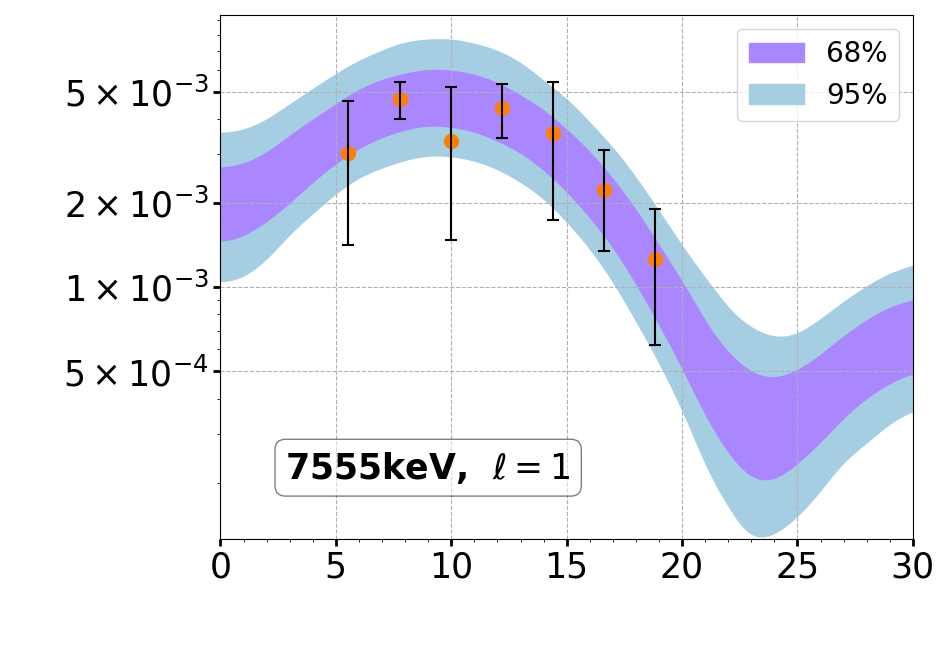}
         \label{fig:7555}
     \end{subfigure}
     \begin{subfigure}[l]{0.3\textwidth}
         \centering
         \includegraphics[ width=\textwidth,trim={40 50 0 0}, clip]{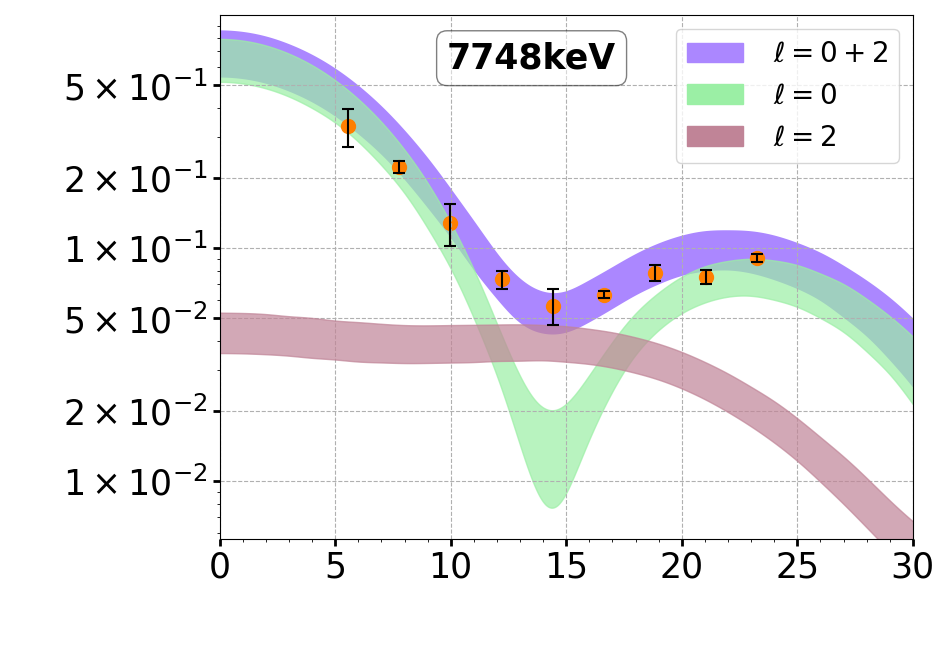}
         \label{fig:7748_twol}
     \end{subfigure}
     \begin{subfigure}[l]{0.3\textwidth}
         \centering
         \includegraphics[ width=\textwidth,trim={40 50 0 0}, clip]{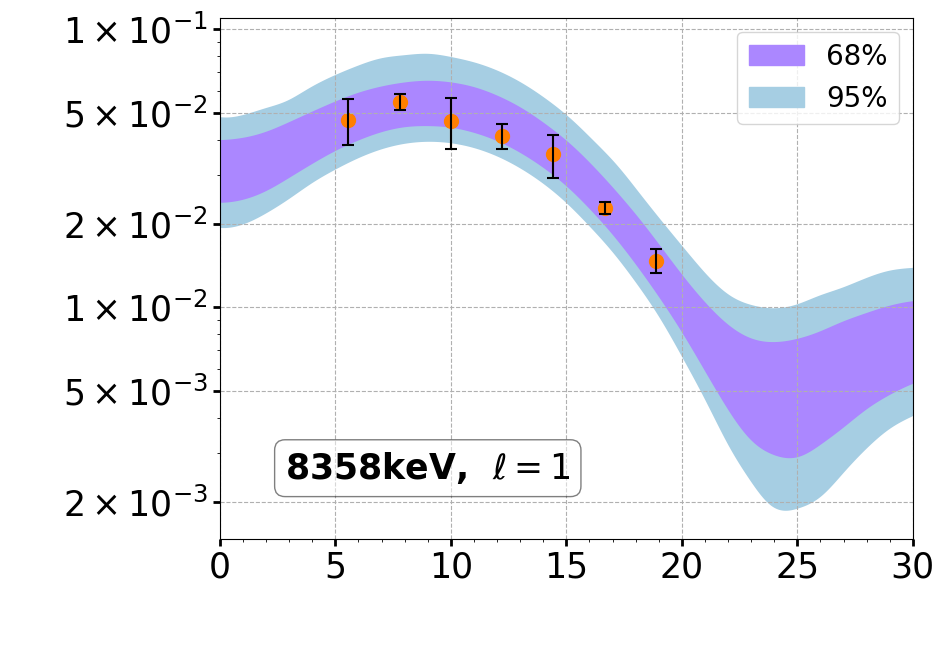}
         \label{fig:8358}
     \end{subfigure}
     \begin{subfigure}[l]{0.3\textwidth}
         \centering
         \includegraphics[ width=\textwidth,trim={40 50 0 0}, clip]{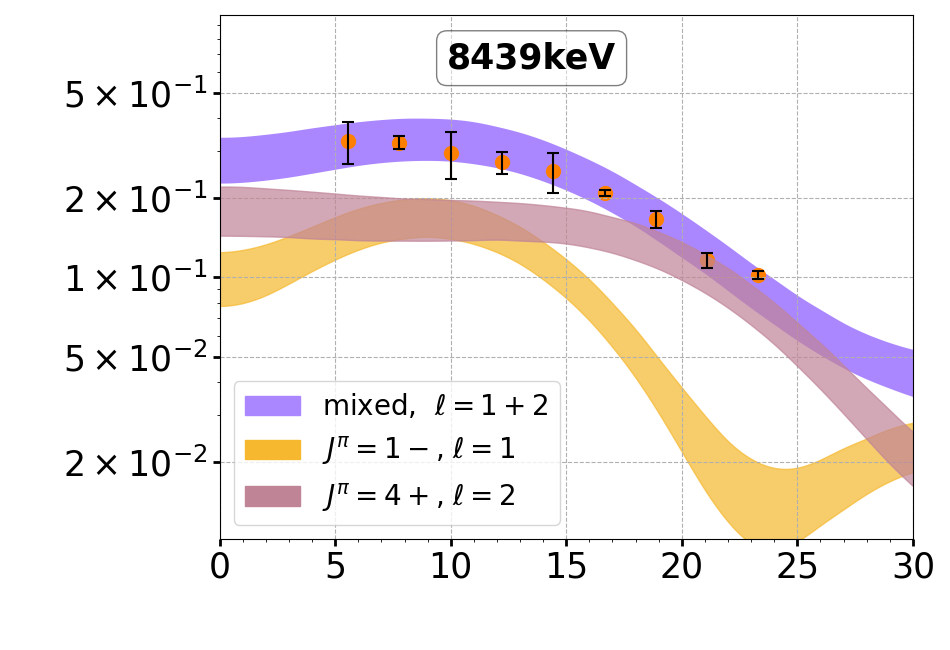}
         \label{fig:8439}
     \end{subfigure}
     \begin{subfigure}[l]{0.3\textwidth}
         \centering
         \includegraphics[ width=\textwidth,trim={40 50 0 0}, clip]{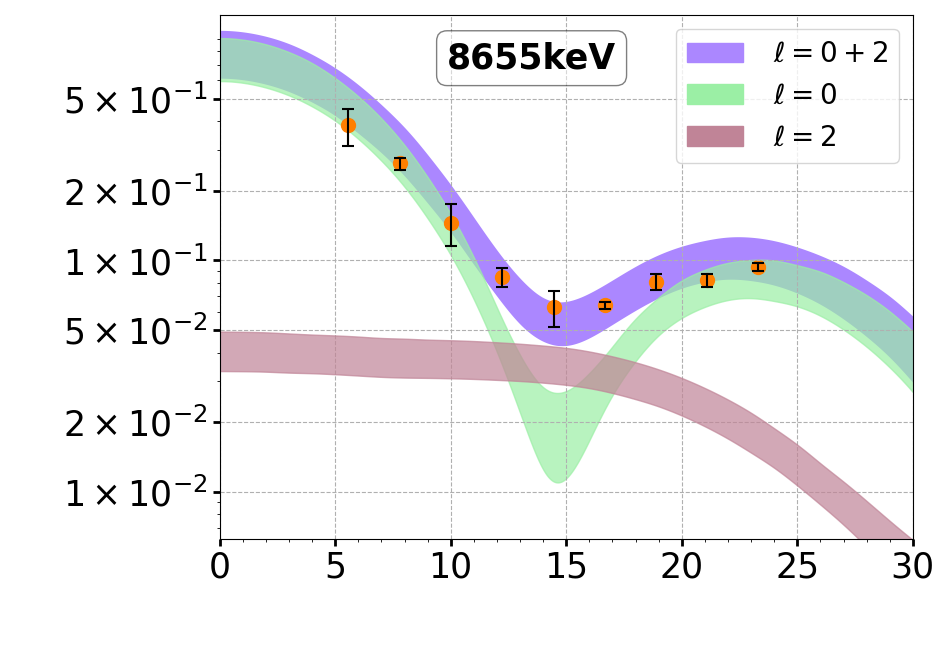}
         \label{fig:8655_twol}
     \end{subfigure}
     \begin{subfigure}[l]{0.3\textwidth}
         \centering
         \includegraphics[ width=\textwidth,trim={40 50 0 0}, clip]{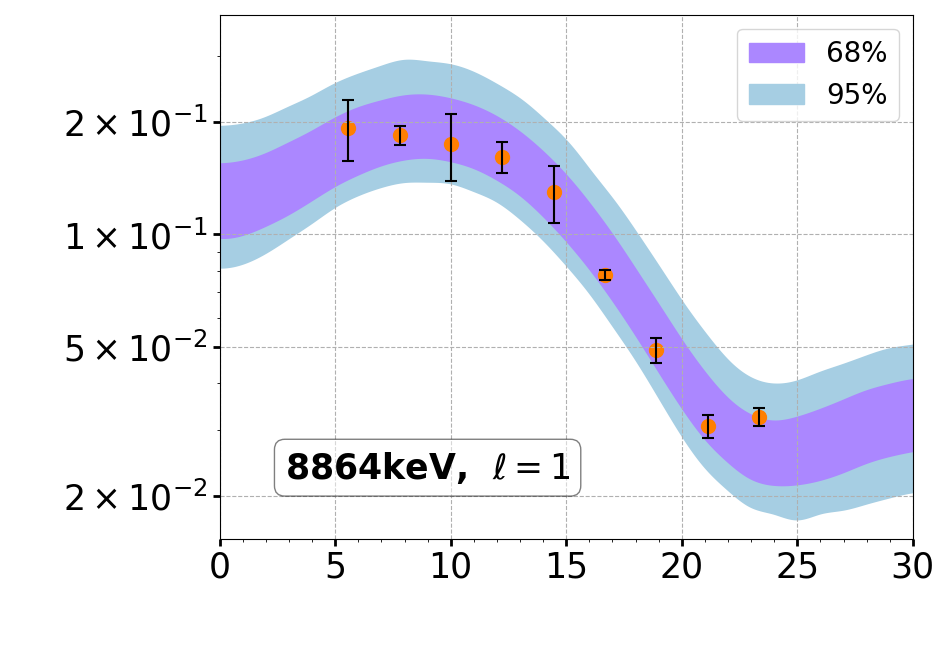}
         \label{fig:8864}
     \end{subfigure}
     \begin{subfigure}[l]{0.3\textwidth}
         \centering
         \includegraphics[ width=\textwidth,trim={40 50 0 0}, clip]{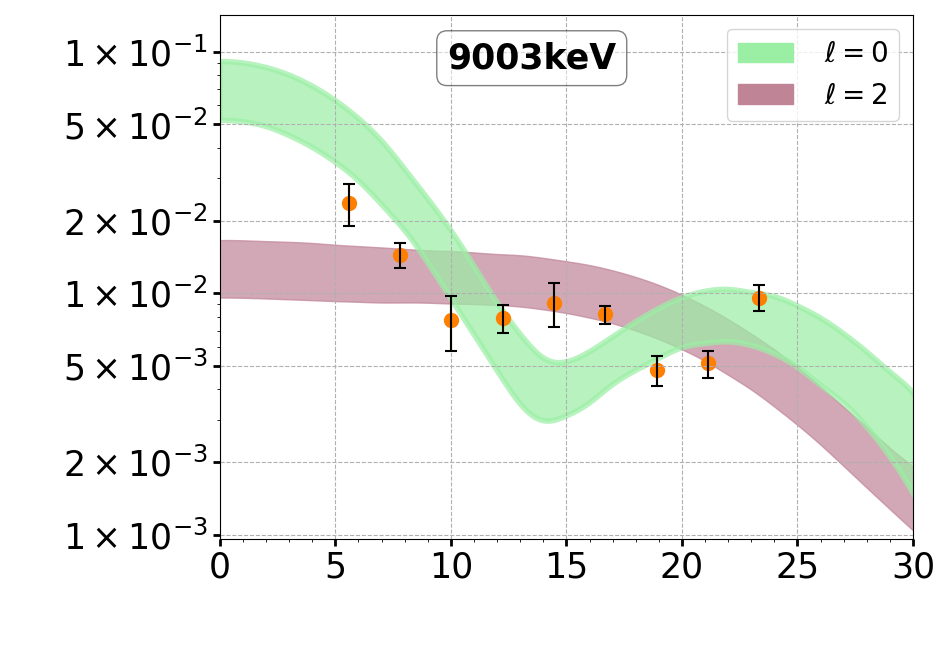}
         \label{fig:9003}
     \end{subfigure}
     \begin{subfigure}[l]{0.3\textwidth}
         \centering
         \includegraphics[ width=\textwidth,trim={40 50 0 0}, clip]{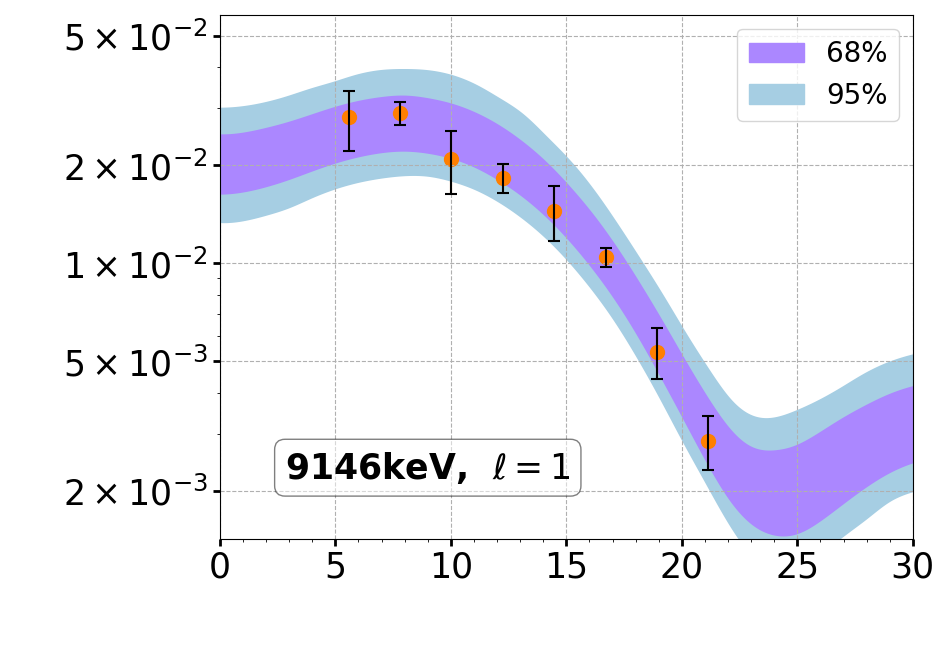}
         \label{fig:9146}
     \end{subfigure}
     \begin{subfigure}[l]{0.3\textwidth}
         \centering
         \includegraphics[ width=\textwidth,trim={40 50 0 0}, clip]{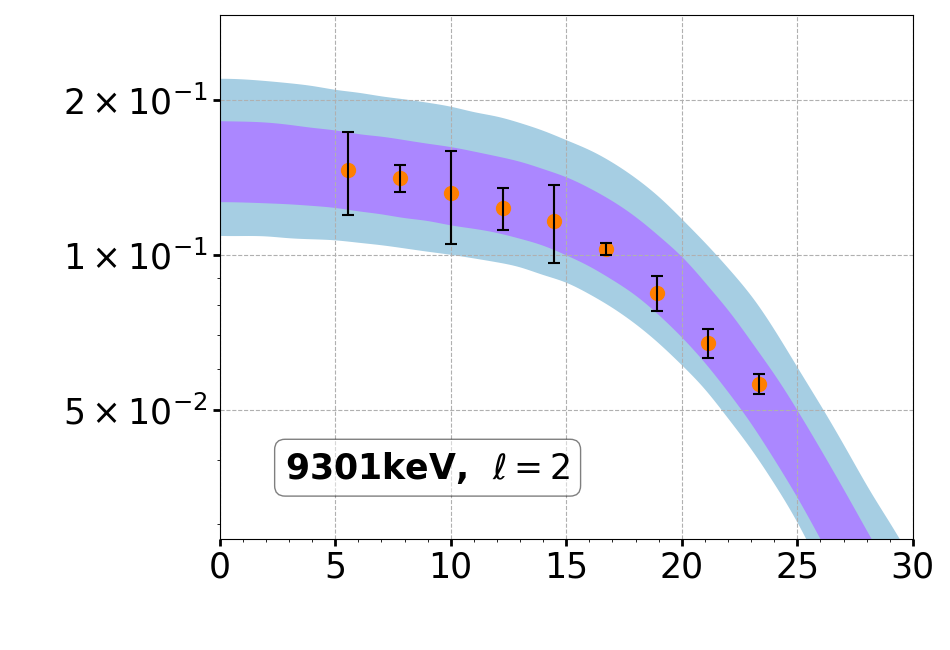}
         \label{fig:9301}
     \end{subfigure}
    % % Add global Y-axis label using \put and \rotatebox
    % \put(10, 5){\rotatebox{90}{\makebox[0pt][l]{\textbf{$\sigma$ (arb.units)}}}}
     \begin{subfigure}[l]{0.3\textwidth}
         \centering
         \includegraphics[ width=\textwidth,trim={40 50 0 0}, clip]{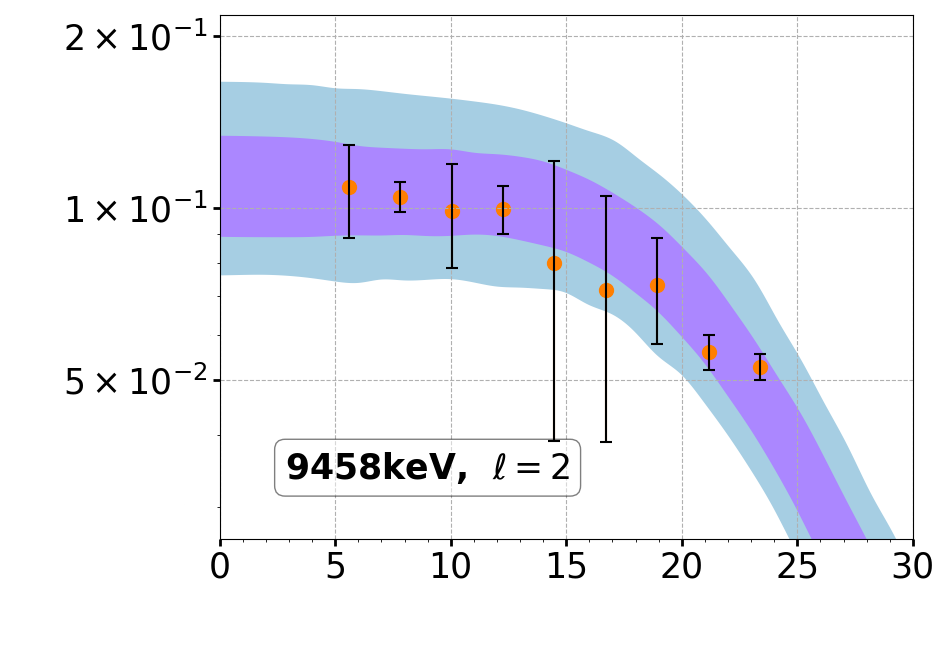}
         \label{fig:9458}
     \end{subfigure}
     \begin{subfigure}[l]{0.3\textwidth}
         \centering
         \includegraphics[ width=\textwidth,trim={40 50 0 0}, clip]{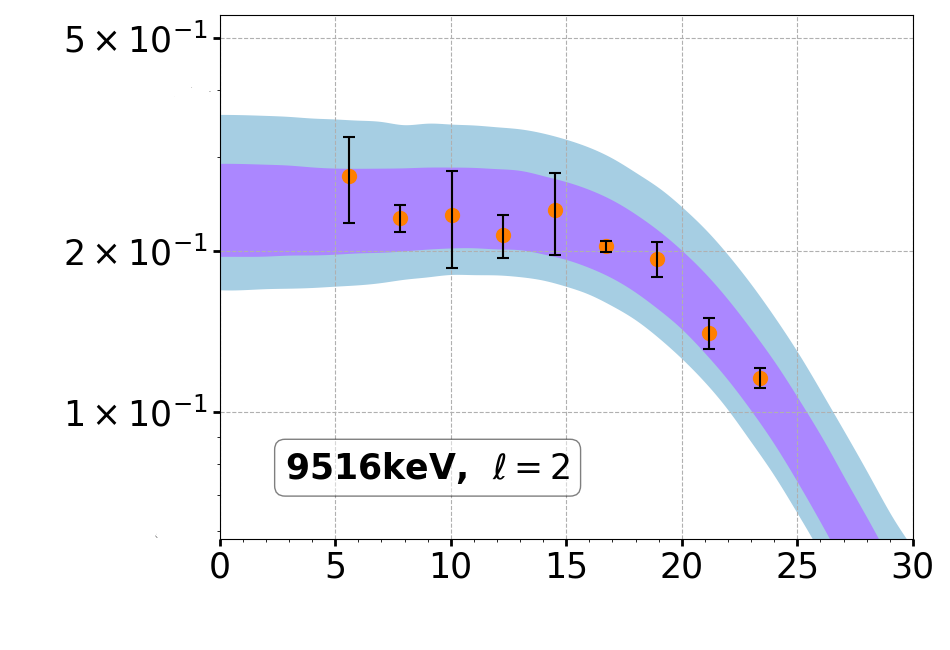}
         \label{fig:9516}
     \end{subfigure}
     \begin{subfigure}[l]{0.3\textwidth}
         \centering
         \includegraphics[ width=\textwidth,trim={40 50 0 0}, clip]{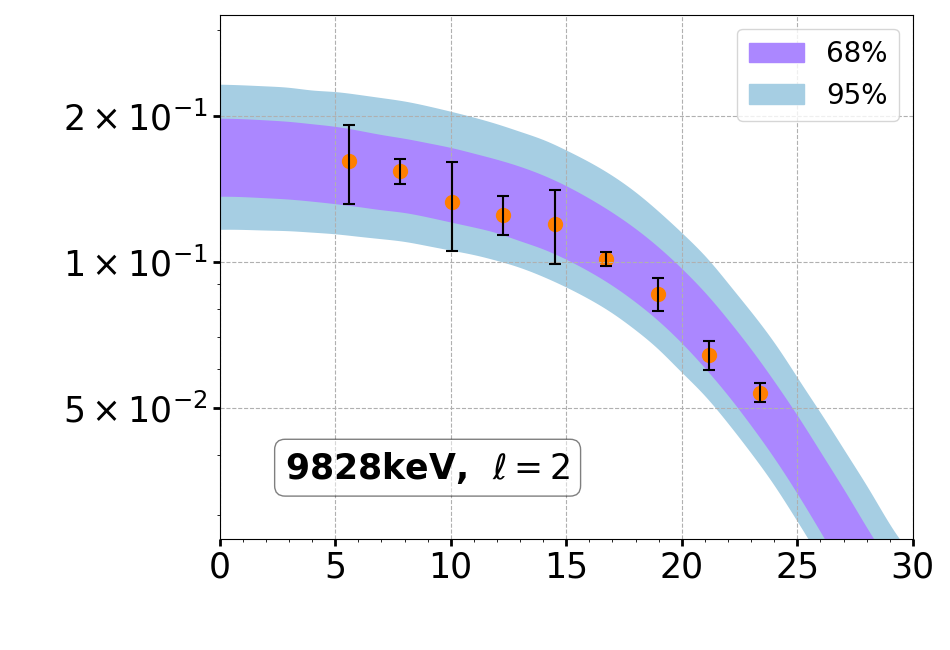}
         \label{fig:9828}
     \end{subfigure}
          \begin{subfigure}[l]{0.3\textwidth}
         \centering
         \includegraphics[ width=\textwidth,trim={40 50 0 0}, clip]{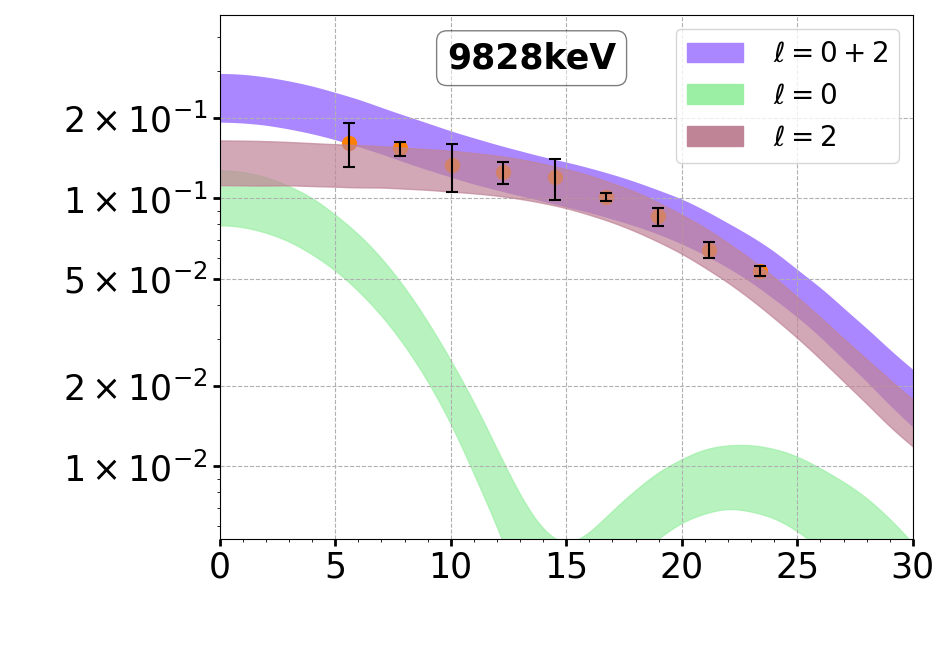}
         \label{fig:9828_twol}
     \end{subfigure}
     \begin{subfigure}[l]{0.3\textwidth}
         \centering
         \includegraphics[ width=\textwidth,trim={40 50 0 0}, clip]{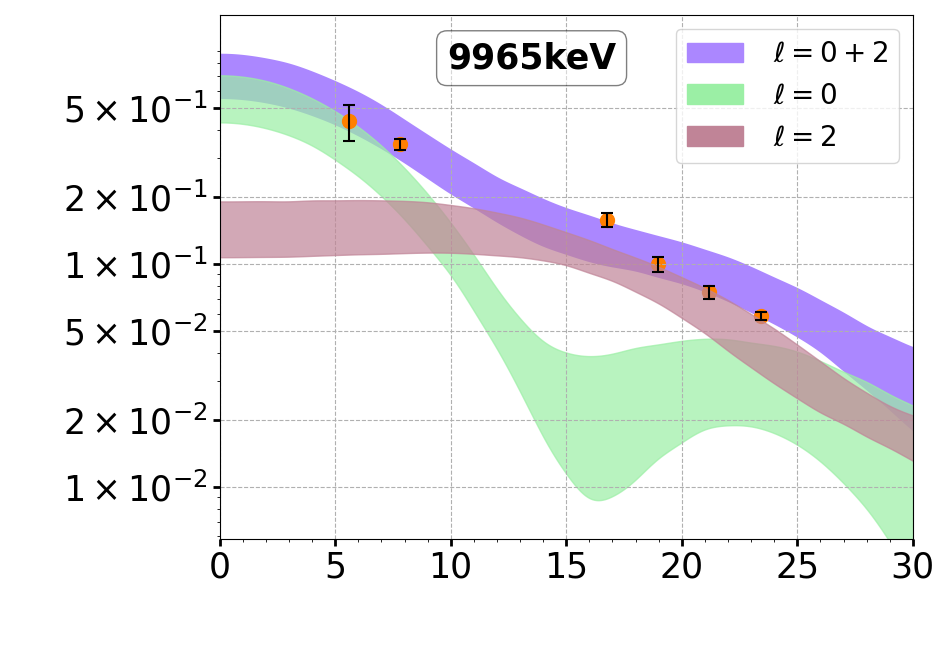}
         \label{fig:9965}
     \end{subfigure}
     \begin{subfigure}[l]{0.3\textwidth}
         \centering
         \includegraphics[ width=\textwidth,trim={40 50 0 0}, clip]{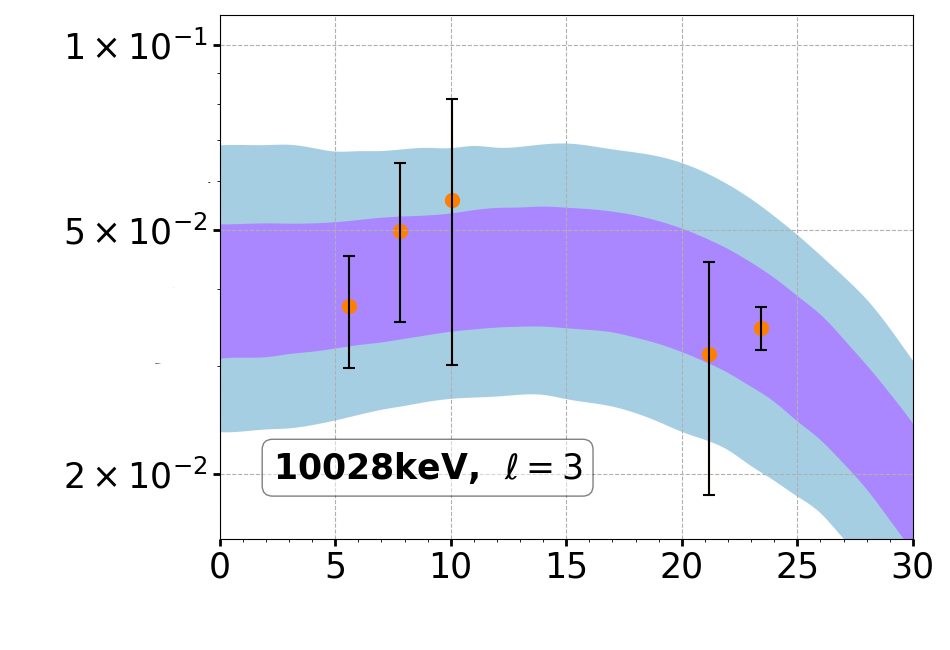}
         \label{fig:10028}
     \end{subfigure}
     \begin{subfigure}[l]{0.3\textwidth}
         \centering
         \includegraphics[ width=\textwidth,trim={40 50 0 0}, clip]{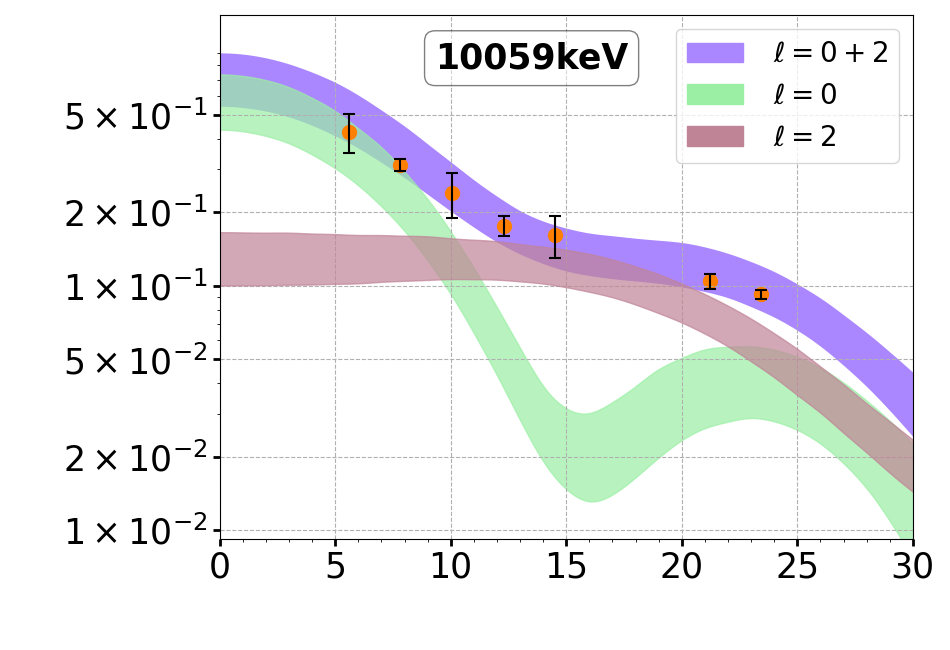}
         \label{fig:10059_twol}
     \end{subfigure}
     \begin{subfigure}[l]{0.3\textwidth}
         \centering
         \includegraphics[ width=\textwidth,trim={40 50 0 0}, clip]{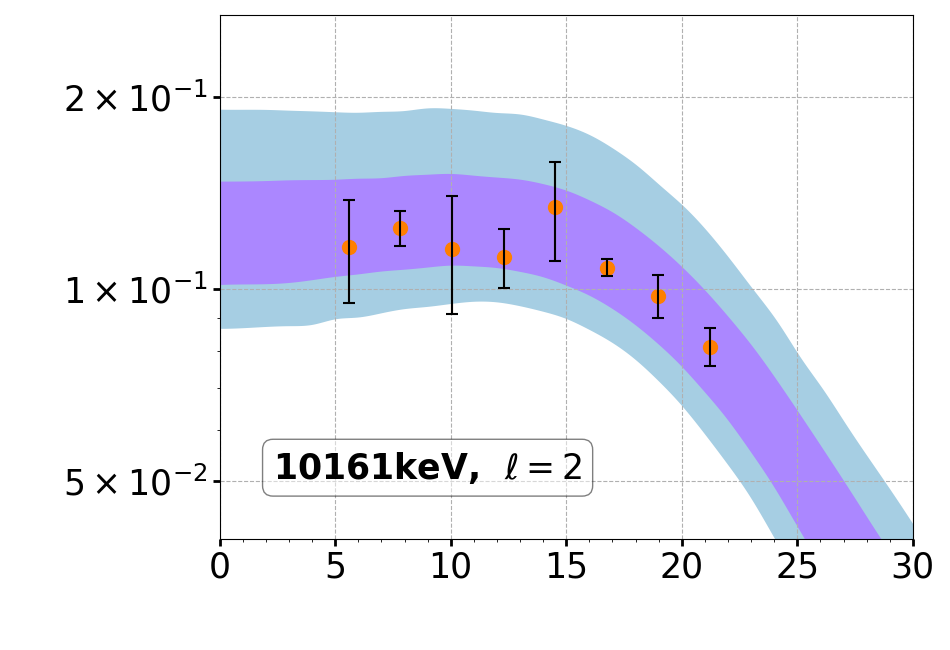}
         \label{fig:10161}
     \end{subfigure}
    \end{minipage}
     \vspace{0.3cm}
        \makebox[\textwidth][c]{\textbf{\normalsize $\theta_{c.m.}$  (deg)}}
        \vspace{-0.6cm}
     % Add global X-axis label using \put
    % \put(-0.45\textwidth,-55){\makebox(0,0){\textbf{$\theta_{c.m.}  (deg)$}}}
        \caption{$^{23}\text{Na}(^3\text{He},d)$ cross sections and MCMC results for all the states}
        \label{fig:MCMC results}
\end{figure*}

\begin{figure*}\ContinuedFloat
     \centering
  \begin{minipage}[c]{0.035\textwidth}
    \centering
    \rotatebox{90}{\textbf{\normalsize $\sigma$ (arb.units)}}
      \end{minipage}%
        \hspace*{-0.4cm}  % bring subfigures closer
  \begin{minipage}[c]{0.96\textwidth}
    \centering
     \begin{subfigure}[l]{0.3\textwidth}
         \centering
         \includegraphics[ width=\textwidth,trim={40 50 0 0}, clip]{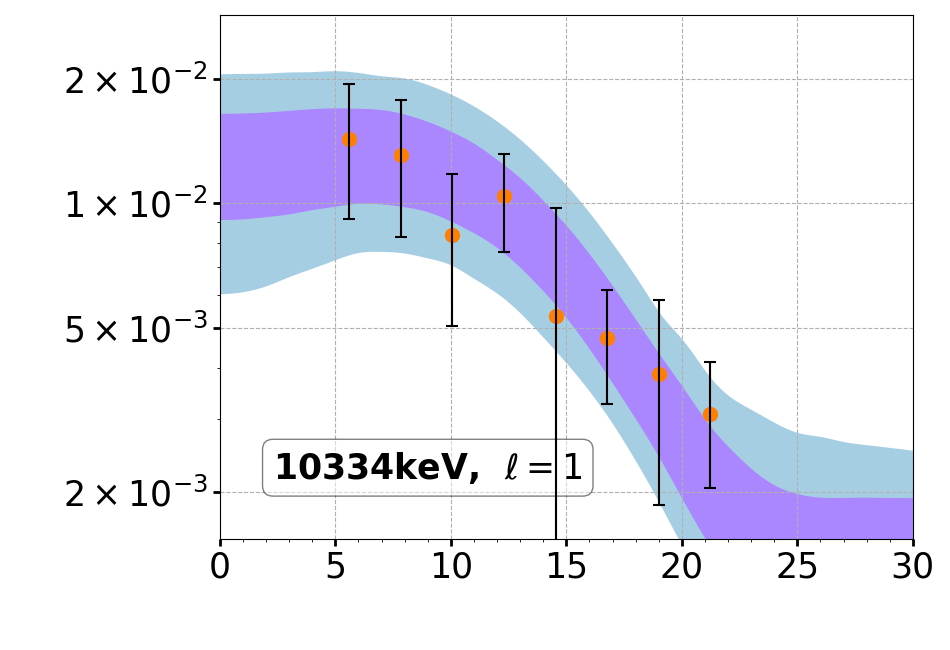}
         \label{fig:10334}
     \end{subfigure}
     \begin{subfigure}[l]{0.3\textwidth}
         \centering
         \includegraphics[ width=\textwidth,trim={40 50 0 0}, clip]{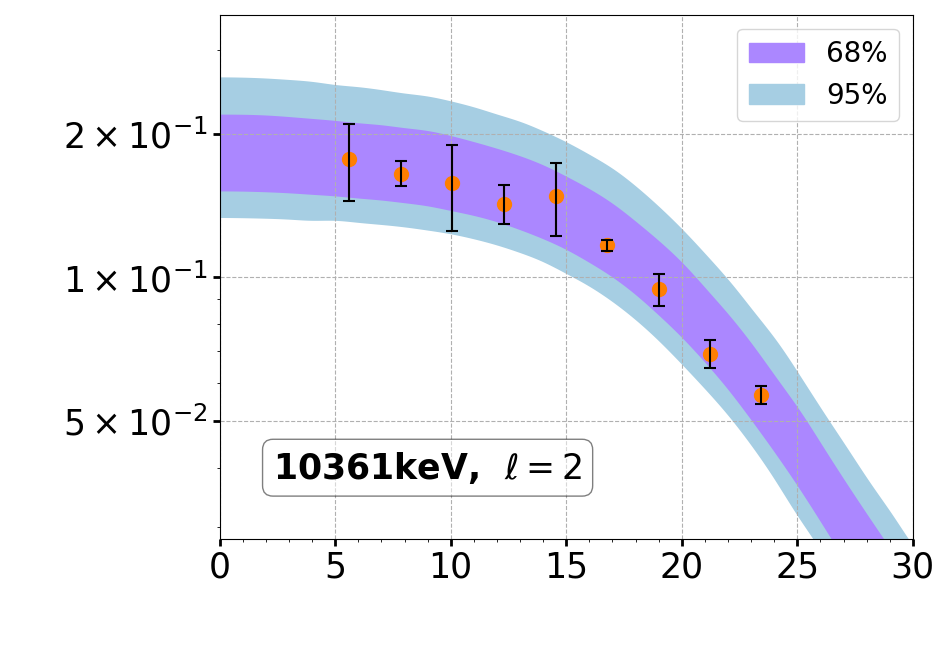}
         \label{fig:10361}
     \end{subfigure}
     \begin{subfigure}[l]{0.3\textwidth}
         \centering
         \includegraphics[ width=\textwidth,trim={40 50 0 0}, clip]{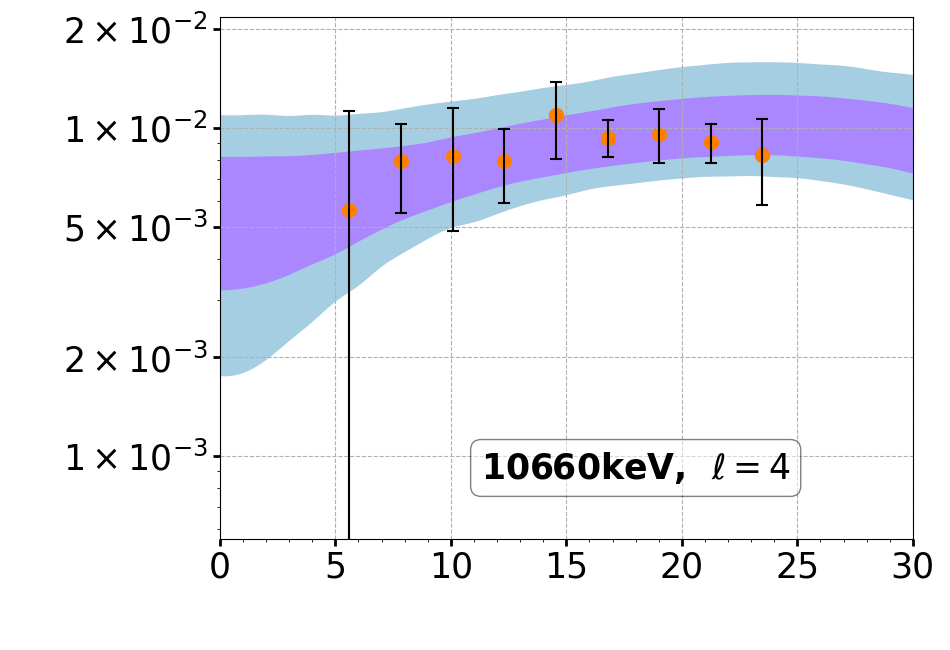}
         \label{fig:10660}
     \end{subfigure}
          \begin{subfigure}[l]{0.3\textwidth}
         \centering
         \includegraphics[ width=\textwidth,trim={40 50 0 0}, clip]{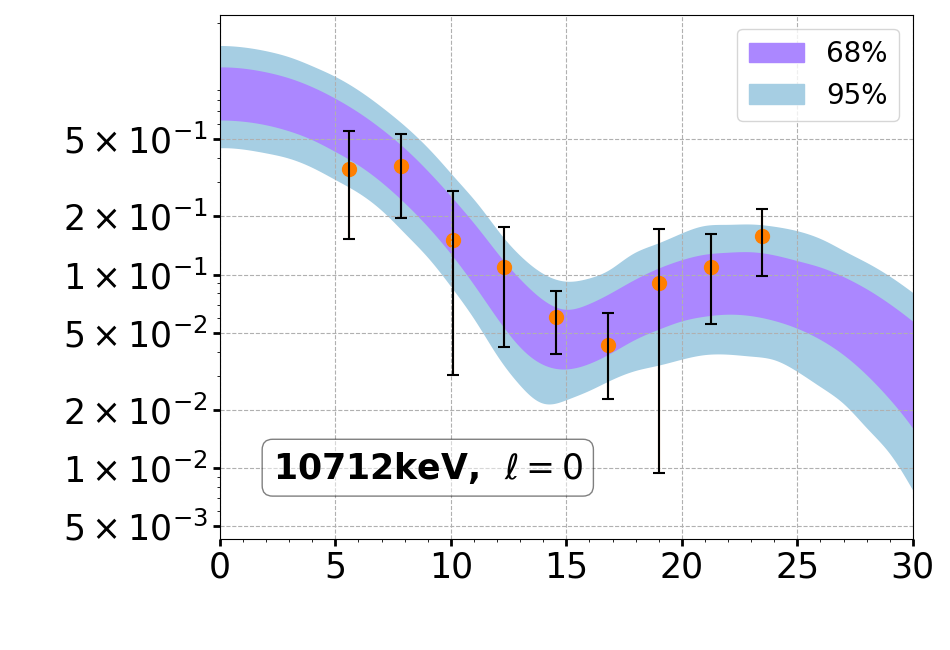}
         \label{fig:10712}
     \end{subfigure}
          \begin{subfigure}[l]{0.3\textwidth}
         \centering
         \includegraphics[ width=\textwidth,trim={40 50 0 0}, clip]{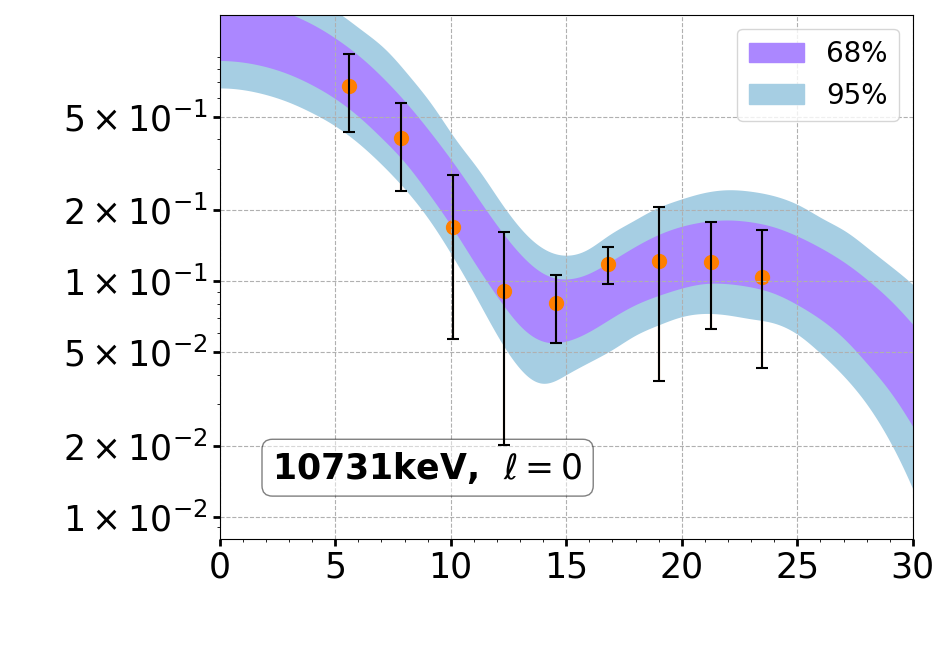}
         \label{fig:10731}
     \end{subfigure}
     \begin{subfigure}[l]{0.3\textwidth}
         \centering
         \includegraphics[ width=\textwidth,trim={40 50 0 0}, clip]{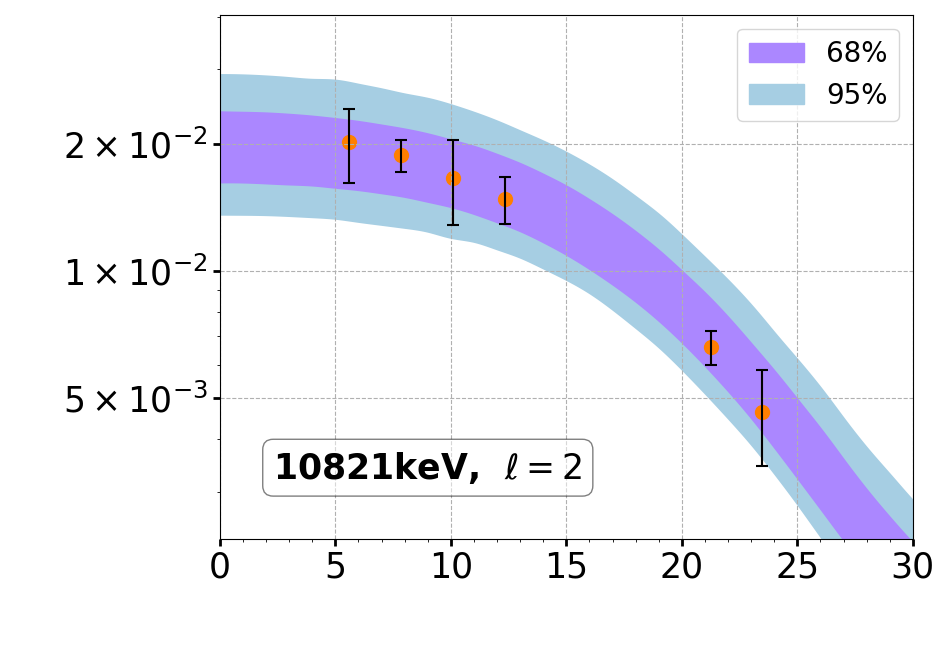}
         \label{fig:10821}
     \end{subfigure}
     \begin{subfigure}[l]{0.3\textwidth}
         \centering
         \includegraphics[ width=\textwidth,trim={40 50 0 0}, clip]{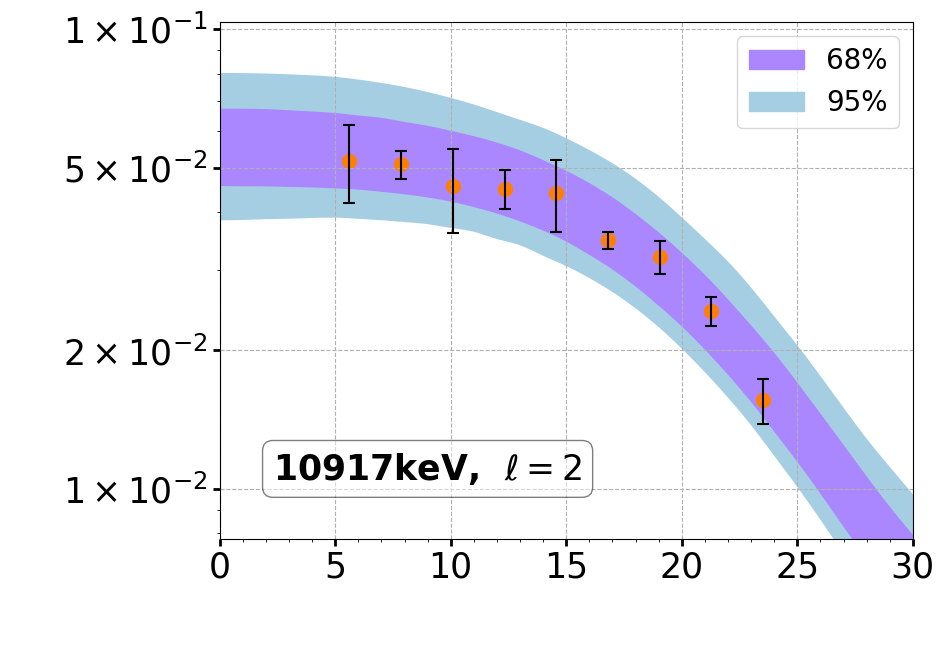}
         \label{fig:10917}
     \end{subfigure}
      \begin{subfigure}[l]{0.3\textwidth}
         \centering
         \includegraphics[ width=\textwidth,trim={40 50 0 0}, clip]{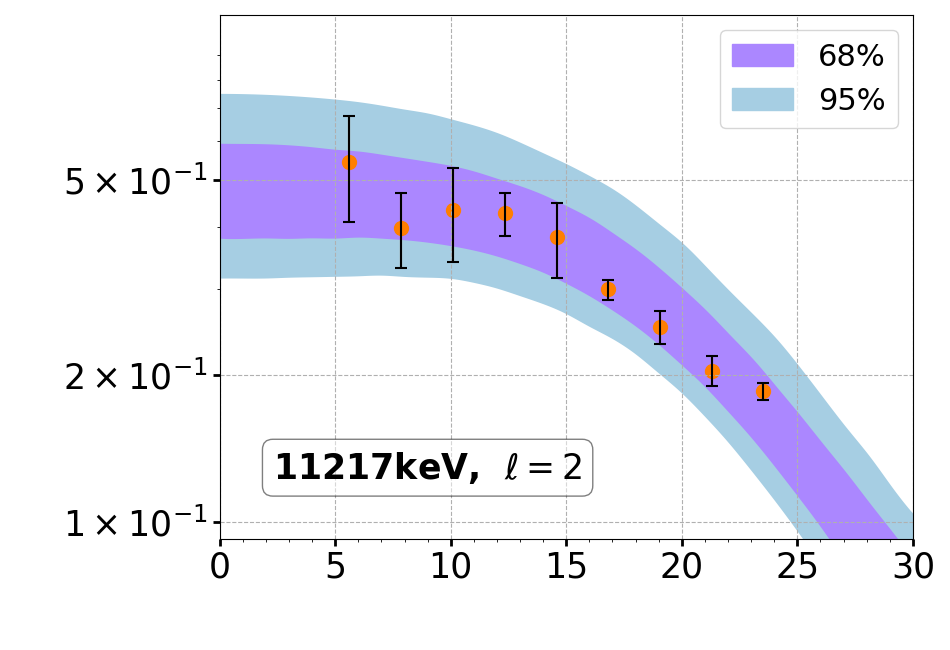}
         \label{fig:11217}
     \end{subfigure}
     \begin{subfigure}[l]{0.3\textwidth}
         \centering
         \includegraphics[ width=\textwidth,trim={40 50 0 0}, clip]{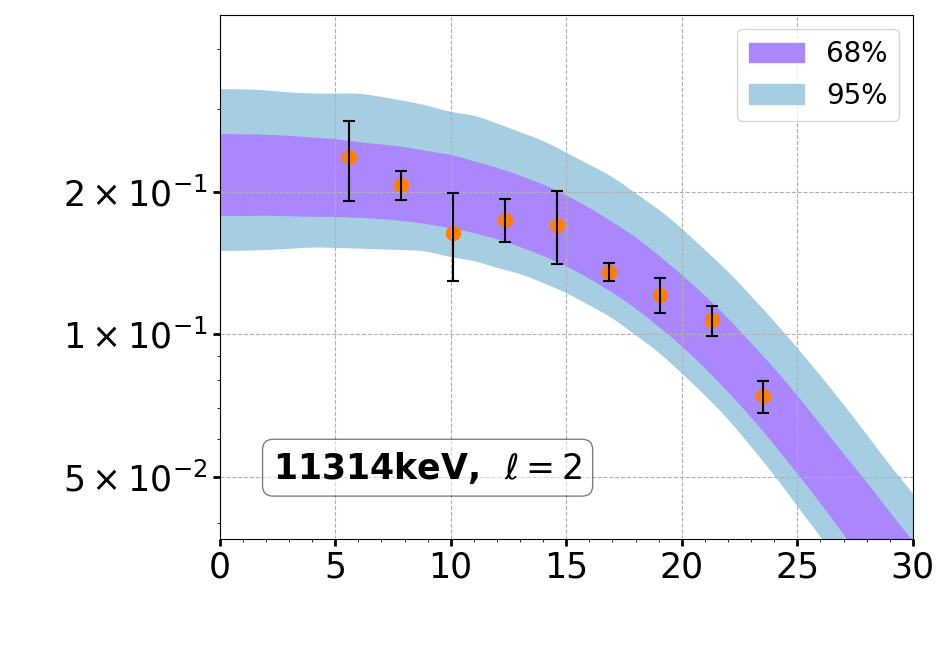}
         \label{fig:11314}
     \end{subfigure}
     \begin{subfigure}[l]{0.3\textwidth}
         \centering
         \includegraphics[ width=\textwidth,trim={40 50 0 0}, clip]{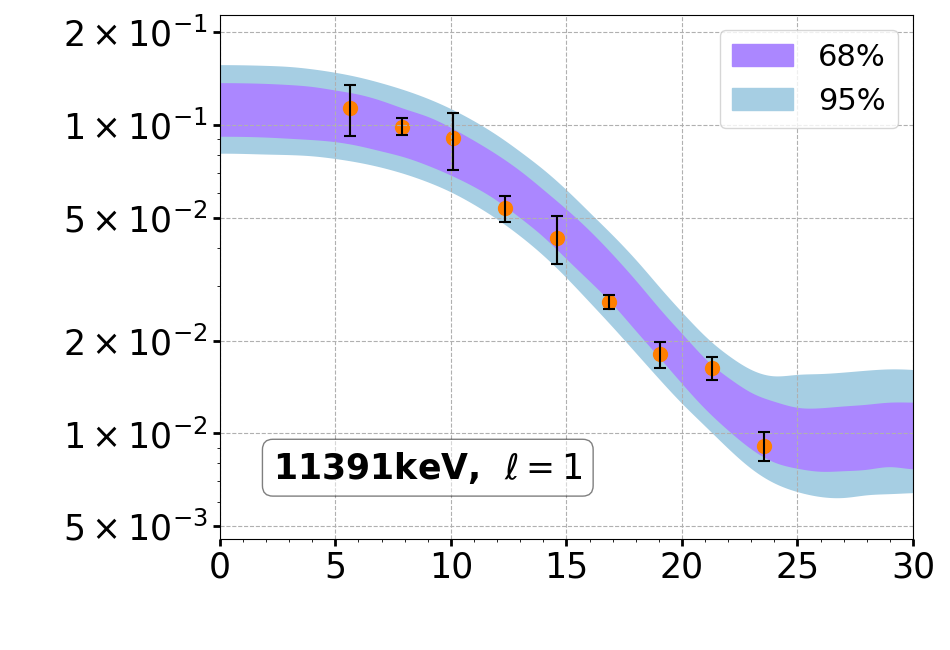}
         \label{fig:11391}
     \end{subfigure}
%              \caption{$^{23}\text{Na}(^3\text{He},d)$ cross sections and MCMC results for all the states}
%         \label{fig:MCMC results}
% \end{figure*}
% \begin{figure*}\ContinuedFloat
%      \centering
     \begin{subfigure}[l]{0.3\textwidth}
         \centering
         \includegraphics[ width=\textwidth,trim={40 50 0 0}, clip]{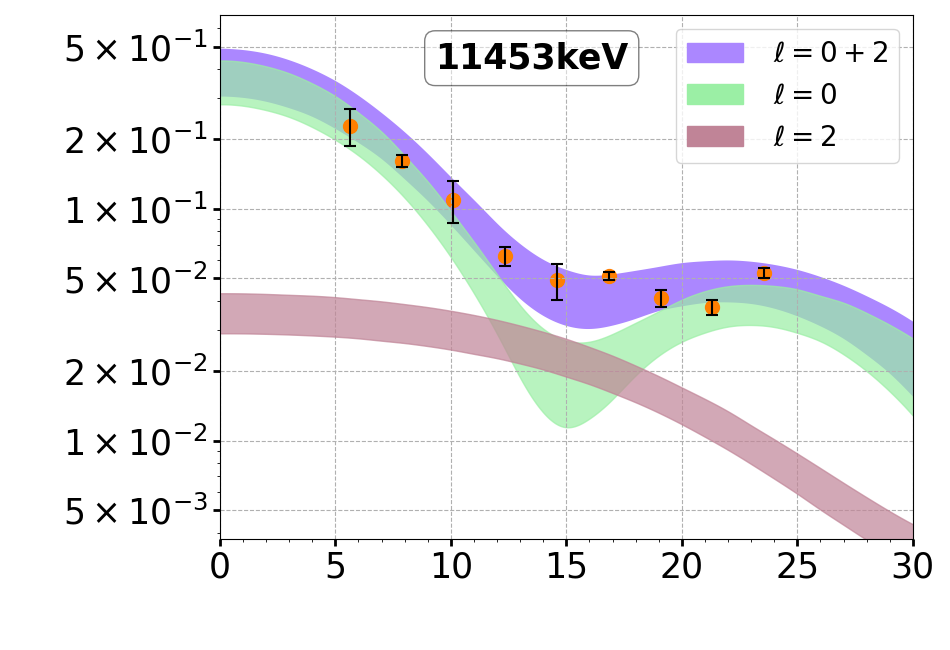}
         \label{fig:11453}
     \end{subfigure}
     \begin{subfigure}[l]{0.3\textwidth}
         \centering
         \includegraphics[ width=\textwidth,trim={40 50 0 0}, clip]{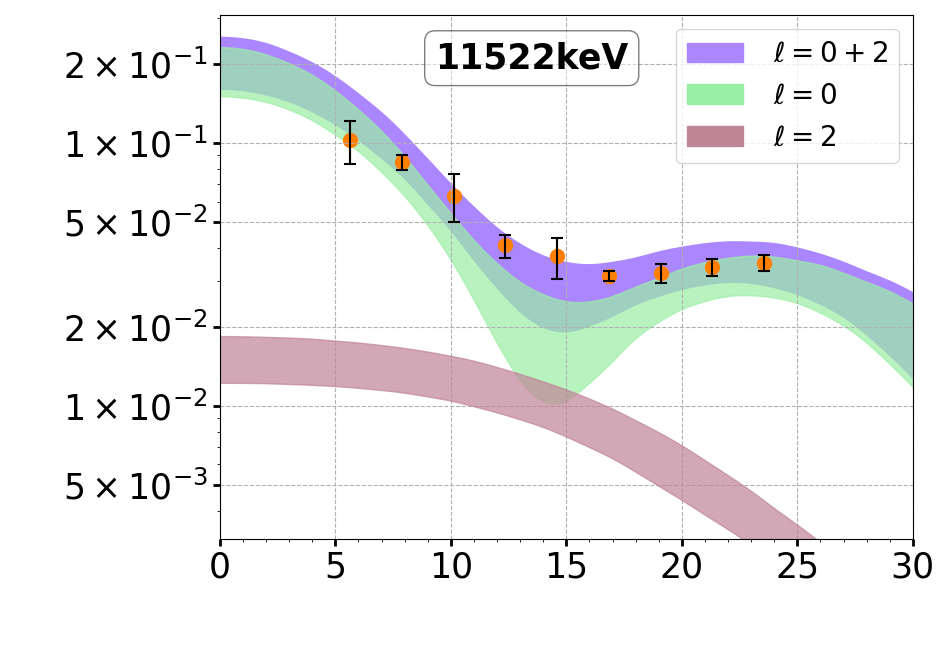}
         \label{fig:11453}
     \end{subfigure}
     \begin{subfigure}[l]{0.3\textwidth}
         \centering
         \includegraphics[ width=\textwidth,trim={40 50 0 0}, clip]{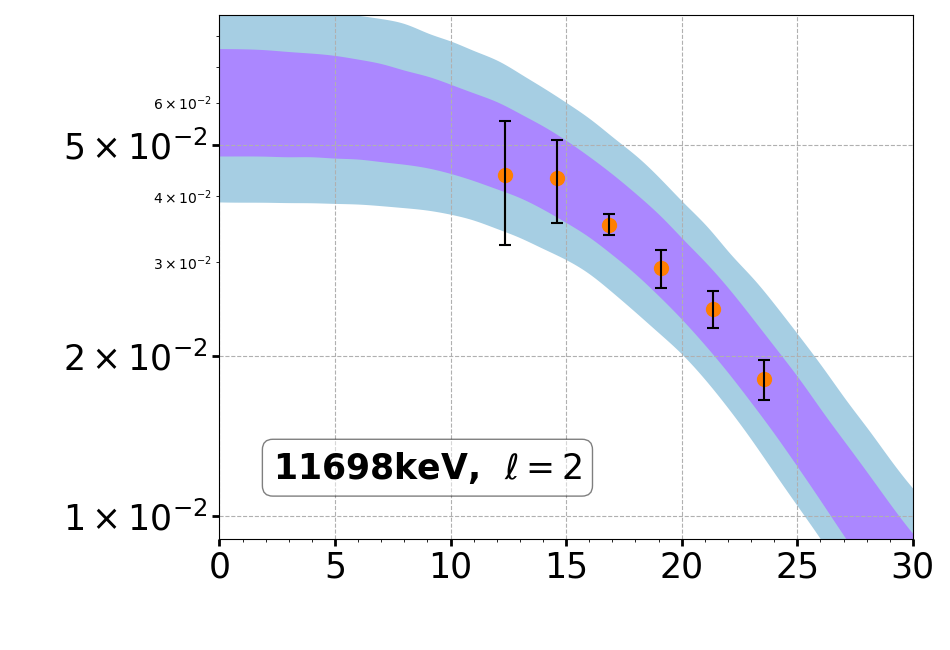}
         \label{fig:11698}
     \end{subfigure}
     \begin{subfigure}[l]{0.3\textwidth}
         \centering
         \includegraphics[ width=\textwidth,trim={40 50 0 0}, clip]{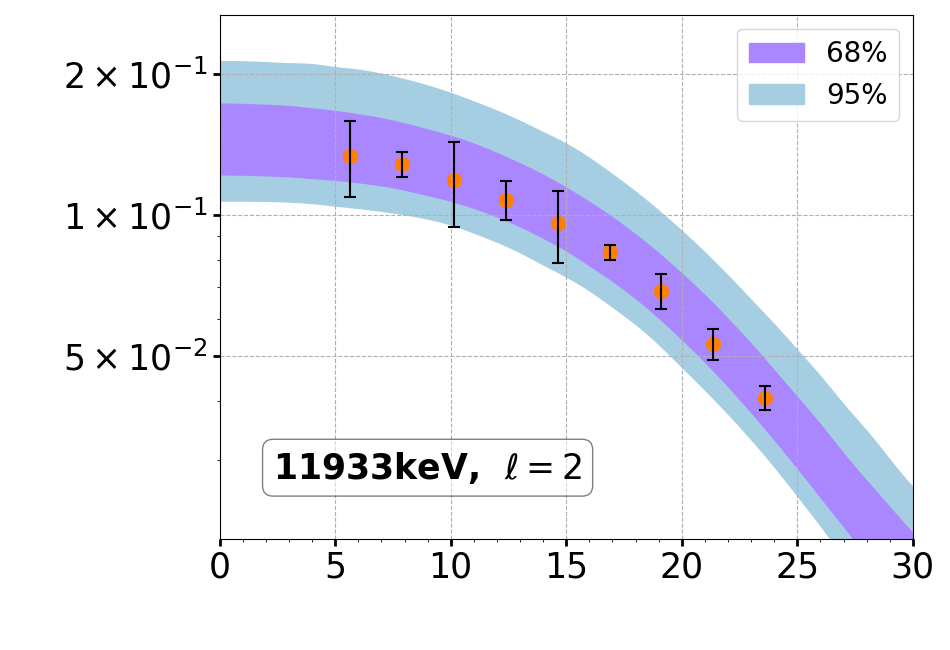}
         \label{fig:11933}
     \end{subfigure}
         \end{minipage}
     \vspace{0.3cm}
        \makebox[\textwidth][c]{\textbf{\normalsize $\theta_{c.m.}$  (deg)}}
        \vspace{-0.6cm}
     \caption{$^{23}\text{Na}(^3\text{He},d)$ cross sections and MCMC results for all the states. The purple and cyan bands gives 68\% and 95\% credential intervals, respectively. For mixed-$\ell$ states, different-$\ell$ components are shown in yellow and red. The y-axis is the yield but not cross section because of the scaling discussed in the text.}
        \label{fig:MCMC results}
\end{figure*}

\subsection{Asymptotic Normalization Coefficients}\label{subsec:ANC_exp}

Often for direct capture and stripping reactions, the reaction is peripheral, and thus dominated by overlap integrals far from the nuclear radius. In this region, the bound state wavefunction is determined by the Coulomb interaction, whose radial solutions, $u_{\ell}(r)$, can be written in terms of the Whittaker function, $W$:
\begin{equation}
    \label{eq:ANC-definition}
    u_{\ell}(r) = C_{\ell} W_{-\eta, l+1/2}(-2kr),
\end{equation}
where $\eta$ is the Sommerfeld parameter, and $C_{\ell}$ is a normalization factor, the Asymptotic Normalization Coefficient (ANC), which describes the strength of the interaction. The ANC depends on details of the bound state and is thus an experimentally-derived quantity. Similarly, a purely theoretical single-particle wavefunction can be calculated by replacing $C_{\ell}$ by the so-called "single-particle ANC", $b_{\ell}$, in Eq. \ref{eq:ANC-definition}. A more thorough discussion of ANCs can be found in e.g., \cite{Mukhamedzhanov1995}.

For the present analysis, the single-particle ANC for states in $^{24}$Mg, $b_{24}$, were extracted from the calculated single-particle wave function and Whittaker ratio for $^{24}$Mg$\rightarrow$$^{23}$Na$+p$ using FRESCO. With our measured $C^2S$, the experimental ANC, $C_{24}$, was determined via:
\begin{equation}
    C_{24}=(C^2S)^{1/2}b_{24}
\end{equation}
The calculation follows Ref.~\cite{Bertone2002}. A column of ANC($C_{24}$) results is provided in Table \ref{tab:summary of C2S}. The negative signs for those values are kept from the original wave functions from $n=2$ orbitals. The uncertainties are half of the relative uncertainties for our measured $C^2S$ since the latter enters in the square root. The ANCs for different orbitals are listed separately in Table. \ref{tab:summary of C2S}, and must be summed to obtain the total ANC for states with multiple $\ell$. 

A comparison between our ANC and Ref.~\cite{Boeltzig22} for several states are shown in Table \ref{tab:ANC_compare}. Most ANCs provided by Ref.~\cite{Boeltzig22} are calculated upper limits for $C_{24}$ and are consistent with our results. For the 10.7 MeV doublet states. Our measured ANC for the two $\ell$ components sum to $4.02$, in good agreement with the value of $3.81(37)$ from Ref.~\cite{Boeltzig22}. 

%More details about the ANC tecnique can be found in \cite{PhysRevC.63.024612}. 
\def\arraystretch{1.2}
\begin{table}[ht]
\caption{\label{tab:ANC_compare}Comparison of ANC values from this work and $(p,\gamma)$ study in Ref.~\cite{Boeltzig22}}
\begin{ruledtabular}
\begin{tabular}{ccccc}
 \multirow{2}{*}{$E_x$(keV)}&\multirow{2}{*}{Orbital}&\multirow{2}{*}{$C^2S$\footnote{$C^2S$ values for $E_x<7$MeV were cited from Garrett \textit{et al.}~\cite{Garrett78} and normalized by a factor of 0.4 to account for the systematic difference. Those values were also used for the direct capture reaction rate calculations.} }&\multicolumn{2}{c}{ANC$(C_{24})$}\\
    &&&This work\footnote{Calculated with $C^2S$ from this table and single-particle ANC($b_{24}$) from FRESCO. }&$(p,\gamma)$ \cite{Boeltzig22}\\
 \hline
0 &$1d_{3/2}$&0.057&3.25&$<$52\\
1369&$2s_{1/2}$&0.0066&-2.55&\multirow{2}{*}{$<$25}\\
 &$1d_{5/2}$&0.2681& 5.98&\\
4123&$1d_{5/2}$&0.0118&0.87&$<$8\\
4238&$2s_{1/2}$&0.0197&-3.30&\multirow{2}{*}{$<$22}\\
 &$1d_{5/2}$&0.0951&2.43& \\
5235&$1d_{5/2}$&0.0556&1.60&$<$11\\
6011&$1d_{5/2}$&0.0036&0.36&$<$11\\
8438($1^-$)&$1p_{1/2}$&0.0639&1.77&\multirow{2}{*}{$<$7}\\
8439($4^+$)&$1d_{5/2}$&0.0551&0.91&\\
\multirow{2}{*}{10731($2^+$)}&$2s_{1/2}$&0.0707&-4.21&\multirow{2}{*}{3.81(37)}\\
 &$1d_{5/2}$&0.0186&0.12& \\
\end{tabular}   
\end{ruledtabular}
\end{table}

\subsection{Direct Capture Reaction Rate}\label{sec:DC}
The reaction rate of $^{23}\text{Na}(p,\gamma)$ is dominated by direct capture at $T\leq50$ MK. For each individual $^{24}$Mg state below the proton threshold $(S_p=11.69\text{MeV})$, the direct capture astrophysical S factor is determined by:  
\begin{align}
(2J_f+1)S_{exp}=(2J_f+1)C^2S\times S_{calc}
\label{eq:S-factor}.
\end{align}

Where $S_{calc}$ is the theoretical S factor calculated assuming pure single-particle states. The \texttt{dircap} code from Christian Illiadis\cite{16Opg} was used to calculate $S_{calc}$. To avoid introducing inconsistencies in the bound state optical potential, the same bound state parameters were used in both direct capture and DWBA calculations. For all $\ell=1$ transitions, the $C^2S$ values for $n\ell=1p$ were used. For $E_x>7$MeV, the recommanded $(2J_f+1)C^2S$ values listed in Table \ref{tab:summary of C2S} are used. For $0\leq E_x<7.3$MeV, which is out of the range of our data, $(2J_f+1)C^2S$ values from Garrett \textit{et al.}~\cite{Garrett78} were incorporated. However, given that their extracted spectroscopic factors are consistently larger than our results and those of Ref.~\cite{Boeltzig22}, we scale the Garrett \textit{et al.} values with a factor of 0.4. Those normalized $C^2S$ values are listed in Table. \ref{tab:ANC_compare}. The total reaction rate is reduced by 15\% as a result of this scaling procedure for states below 7.3 MeV. 
% we made two separate calculations with $(2J_f+1)C^2S$ values incorporated from two different sources: 1. Garrett \textit{et al.}~\cite{Garrett78}. 2. Tang \textit{et al.}~\cite{Tang}. We used the former for further reaction rate calculations. Using the later would increase the total DC S factor and the DC reaction rates by 10\% in the range of interest. The cited $(2J_f+1)C^2S$ are normalized to the average ratios presented in Fig. \ref{fig:C2S_Ratio} for consistency. 

The ground state was not included in Garrett \textit{et al.}~\cite{Garrett78}. Large discrepancies exist among literature for the measured $(2J_f+1)C^2S$ of the ground state: 0.025 from Ref.~\cite{Bearse}, 0.14 from Ref.~\cite{Fuchs} and 0.50 from Ref.~\cite{Tang}. The value from Ref.~\cite{Fuchs} were adopted among those since their measured $(2J_f+1)C^2S$ values for the other states are systematically consistent with Ref.~\cite{Garrett78}. The adopted value is also normalized by a factor of 0.4 as discussed above.

A study by Iliadis \textit{et al.} Ref.~\cite{SF_DC} showed that compared with traditionally adopted hard sphere(HS) potential, DC cross sections with a zero scattering(ZS) potential brings better consistency between transfer reaction and $(p,\gamma)$ results. In this work, we present results from both models.

For each state, the direct capture cross sections and astrophysical S factor $S_{\text{exp}}$ are calculated for $E < 1.5$ MeV in 10 keV steps. The total DC S factor, $S(E)$, is then determined by incoherently summing $S_{\text{exp}}$ over all the states at each energy step. A quadratic fit for $S(E)$ in the range of $E_{c.m.}<1.5$ MeV  gives:
\begin{align}
S_{\text{ZS}}(E)=0.0130-0.00179\cdot E+0.000809\cdot E^2 (\text{MeV}\cdot b)\\
S_{\text{HS}}(E)=0.0115-0.00271\cdot E+0.000644\cdot E^2 (\text{MeV}\cdot b)
\label{eq:S(E)}.
\end{align}

The uncertainties of $S(E)$ are estimated to be 43\%, plus 2\% from the quadratic fits in the range.

\begin{figure*}[htbp]
     \includegraphics[width=\linewidth]{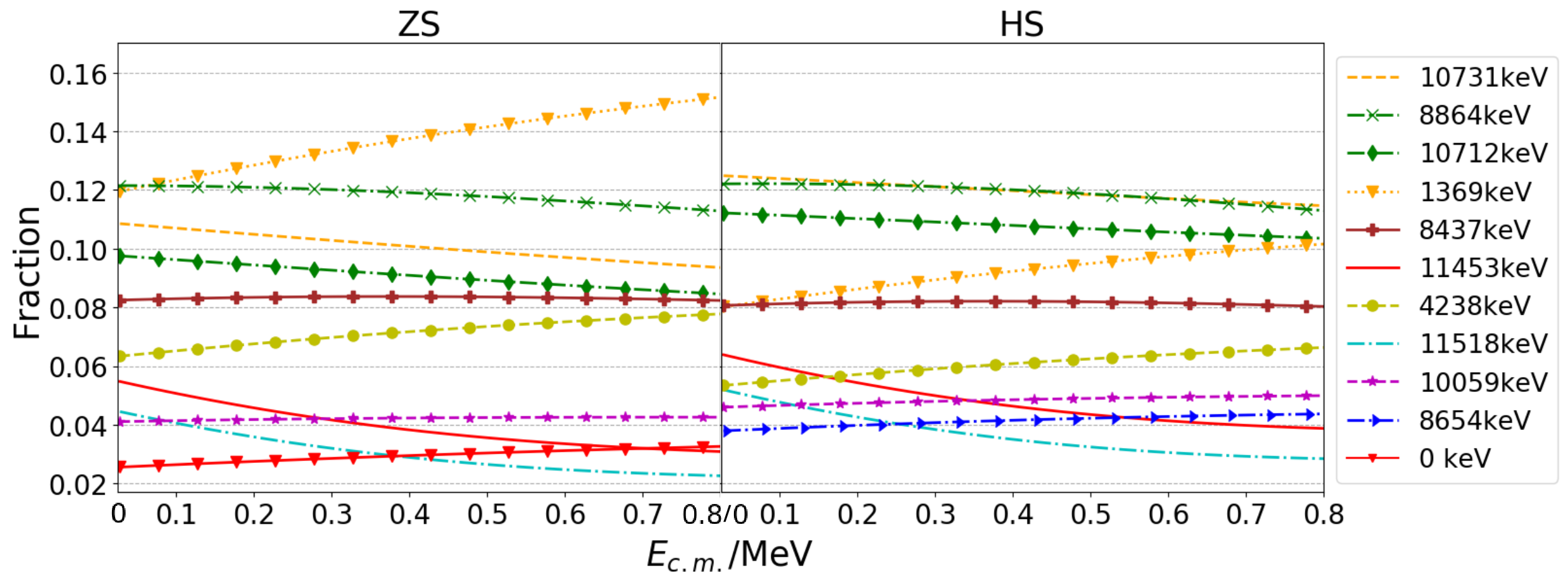}
\caption{\label{fig:S_contributions} Fractional contributions to the total DC S factor with $C^2S$ values from this work. Left: zero scattering potential. Right: hard sphere potential. 10 states with the most contributions are plotted. For $E_x<7$MeV, normalized $C^2S$ values from Garrett were adopted. See the text for more details.}
\end{figure*}

 Fig. \ref{fig:S_contributions}. shows the relative contributions of 10 states that are the most significant to the total direct capture cross section from the two models. States shown in the figure are summations from all the allowed angular momentum $\ell=0,1,2$ and doublet components based on Table \ref{tab:summary of C2S}. The low energy states contribute more to the total cross section in the zero scattering model. In both cases, the 10712 keV and 10731 keV states contribute significantly to the cross section, $\sim23\%$ in total. The other significant contributions all have small $\ell$ components, mostly s-wave transitions. A similar calculation was performed by Ref.~\cite{Boeltzig22} using only $C^2S$ values from Garrett \textit{et al.} Compared with their results, the 8864 keV state's contribution is larger in this work since the $C^2S$ value from $nl=1p$ was adopted. The 8655 keV state contributed less compared with the newly included states and was not included in the figure. Moreover, the 7748 keV state contributes less in this work due to the systematic difference of $C^2S$ between our work and the literature, as shown in Fig.~\ref{fig:C2S_Ratio}. 
\def\arraystretch{1.3}
\begin{table}[ht]
\centering
\caption{\label{tab:Reaction_Rate}%
$^{23}$Na$(p,\gamma)^{24}$Mg Reaction Rate
}
\begin{tabular}{ccccc}
\hline\hline
\multirow{2}{*}{T9(K)}&
\multirow{2}{*}{Low}&
\multirow{2}{*}{Recommended}&
\multirow{2}{*}{High}&
\multirow{2}{*}{f.u.}\\
\\
\colrule
0.010 & 4.77E-33 & 6.72E-33 & 9.70E-33 & 1.41 \\
0.011 & 9.14E-32 & 1.29E-31 & 1.86E-31 & 1.41 \\
0.012 & 1.25E-30 & 1.75E-30 & 2.53E-30 & 1.41 \\
0.013 & 1.29E-29 & 1.81E-29 & 2.61E-29 & 1.41 \\
0.014 & 1.06E-28 & 1.49E-28 & 2.15E-28 & 1.41 \\
0.015 & 7.16E-28 & 1.01E-27 & 1.45E-27 & 1.41 \\
0.016 & 4.12E-27 & 5.80E-27 & 8.36E-27 & 1.41 \\
0.018 & 9.10E-26 & 1.28E-25 & 1.84E-25 & 1.41 \\
0.020 & 1.31E-24 & 1.84E-24 & 2.65E-24 & 1.41 \\
0.025 & 2.71E-22 & 3.82E-22 & 5.50E-22 & 1.41 \\
0.030 & 1.58E-20 & 2.23E-20 & 3.21E-20 & 1.41 \\
0.04 & 6.51E-18 & 8.99E-18 & 1.28E-17 & 1.39 \\
0.05 & 9.35E-16 & 1.42E-15 & 2.41E-15 & 1.61 \\
0.06 & 6.83E-14 & 1.20E-13 & 2.20E-13 & 1.79 \\
0.07 & 2.01E-12 & 3.47E-12 & 6.13E-12 & 1.75 \\
0.08 & 2.87E-11 & 4.61E-11 & 7.69E-11 & 1.65 \\
0.09 & 3.06E-10 & 4.26E-10 & 6.33E-10 & 1.45 \\
0.1 & 3.40E-09 & 4.11E-09 & 5.04E-09 & 1.23 \\
0.2 & 6.49E-03 & 7.63E-03 & 8.94E-03 & 1.18 \\
0.3 & 9.99E-01 & 1.19E+00 & 1.40E+00 & 1.19 \\
0.4 & 1.12E+01 & 1.33E+01 & 1.57E+01 & 1.19 \\
0.5 & 4.48E+01 & 5.33E+01 & 6.26E+01 & 1.19 \\
0.6 & 1.10E+02 & 1.30E+02 & 1.52E+02 & 1.18 \\
0.7 & 2.05E+02 & 2.42E+02 & 2.82E+02 & 1.18 \\
0.8 & 3.31E+02 & 3.86E+02 & 4.47E+02 & 1.17 \\
0.9 & 4.84E+02 & 5.59E+02 & 6.42E+02 & 1.16 \\
1.00 & 6.65E+02 & 7.59E+02 & 8.63E+02 & 1.14 \\
1.25 & 1.24E+03 & 1.38E+03 & 1.54E+03 & 1.11 \\
1.50 & 2.01E+03 & 2.20E+03 & 2.38E+03 & 1.09 \\
1.75 & 2.97E+03 & 3.20E+03 & 3.44E+03 & 1.08 \\
2.0 & 4.10E+03 & 4.40E+03 & 4.69E+03 & 1.07 \\
2.5 & 6.79E+03 & 7.22E+03 & 7.64E+03 & 1.06 \\
3.0 & 9.77E+03 & 1.04E+04 & 1.10E+04 & 1.06 \\
3.5 & 1.28E+04 & 1.35E+04 & 1.43E+04 & 1.06 \\
4.0 & 1.55E+04 & 1.64E+04 & 1.74E+04 & 1.06 \\
5.0 & 2.01E+04 & 2.13E+04 & 2.26E+04 & 1.06 \\
6.0 & 2.32E+04 & 2.46E+04 & 2.61E+04 & 1.06 \\
7.0 & 2.51E+04 & 2.66E+04 & 2.83E+04 & 1.06 \\
8.0 & 2.61E+04 & 2.77E+04 & 2.94E+04 & 1.06 \\
9.0 & 2.65E+04 & 2.81E+04 & 2.99E+04 & 1.06 \\
10.0 & 2.64E+04 & 2.81E+04 & 2.98E+04 & 1.06 \\

\hline\hline
\end{tabular}
\end{table}

\begin{figure}[ht]
\includegraphics[width=70mm]{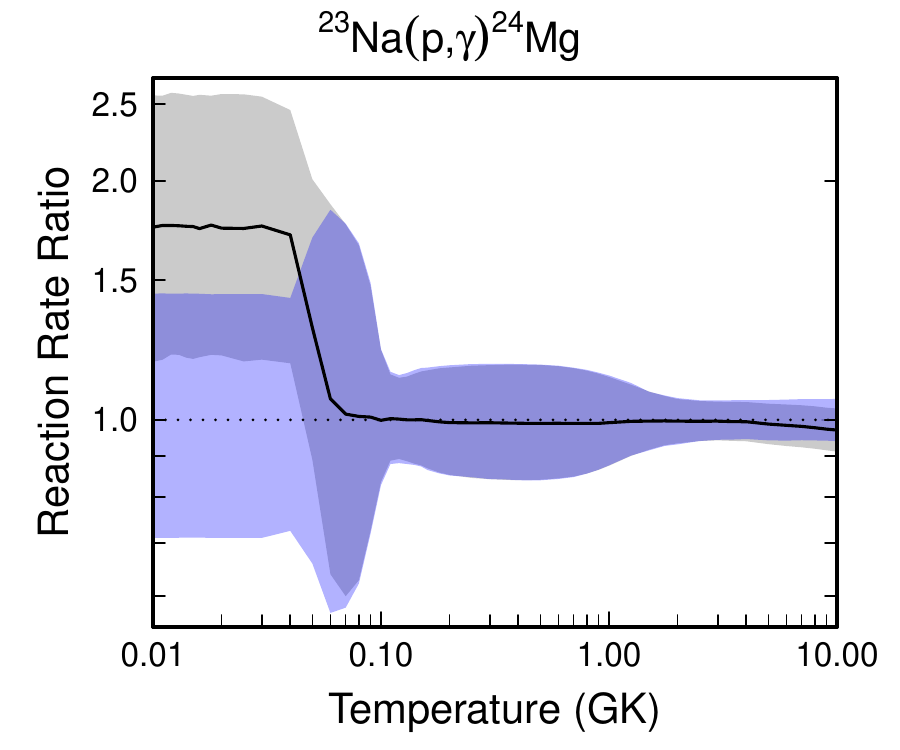}% Here is how to import EPS art
\caption{\label{fig:GraphCompare.pdf} The ratio of $^{23}$Na$(p,\gamma)$ reaction rate between this work(blue) and Ref.~\cite{Marshall2021}(gray). The shaded areas represent $1-\sigma$ uncertainties. The presented reaction rates were calculated based on the ZS potential DC model(see the text). The ratio would start from 2.0 at 0.01GK if HS potential were used instead.}
\end{figure}

New reaction rates for $^{23}$Na$(p,\gamma)$ were subsequently calculated using RatesMC \cite{RatesMC1}, incorporating the updated direct capture S factor (with the zero scattering model) from this work and resonance parameters from Ref.~\cite{Marshall23}. The results are listed in Table \ref{tab:Reaction_Rate}. The associated factor uncertainties are presented as \textit{f.u.} in the table. A comparison with the previous result is shown in Fig.\ref{fig:GraphCompare.pdf}, where the solid line and blue area denotes our present rate relative to Ref.~\cite{Hale04}, which is plotted with dashed line and gray area. The shaded area denotes $1-\sigma$ uncertainty. 

The new reaction rate is smaller than the previous rate by $\sim43\%$ below $T_9 = 0.04$, mainly due to a systematic difference between our extracted $C^2S$ values and the previous adopted Garrett \textit{et al.}~\cite{Garrett78} values, as plotted in Fig. \ref{fig:C2S_Ratio}. We have reproduced the DWBA calculations in Garrett \textit{et al.}~\cite{Garrett78} using their optical model potential parameters and verified that $20\%$ of the difference is explained by our different choice of parameters, but the majority of the difference is unexplained, and is possibly due to experimental effects. Aside from these effects, a small difference ($\sim7\%$) in reaction rates arises from applying the zero-scattering DC model \cite{16Opg} instead of the hard-sphere model. 

% {\color{orange}{Our presented reaction rate has included updates from Ref.~\cite{Marshall2021},\cite{Marshall23} for resonance energies near the threshold. Other measurements on the resonances can be found in Ref.~\cite{Cesaratto2013},\cite{BOELTZIG2019}. Contributions from the broad resonance tails recently presented in Ref.~\cite{Boeltzig22} was not included in our calculations.}\color{red}{ - \textbf{This becomes unnecesary if we just use Marshall. However, should we use Boelzig to compare since he also did a DC calculation? -- I think ours is considered independent/parallel from theirs. Their (p,g) measurement updated the ANC for the 10.7 MeV state (which kind of agrees with our new C2S) and broad resonance tails from 3 unbound states. (Can we include this in RatesMC?) However, they said in the paper that Caleb's new resonance energies (2021) were not included in their calculations. That made things inconsistent. }}} 

Our presented reaction rate has included updates from Refs.~\cite{Marshall2021},\cite{Marshall23} for resonance energies near the threshold. Other measurements on the resonances can be found in Refs.~\cite{Cesaratto2013},\cite{BOELTZIG2019}. Contributions from the broad resonance tails recently presented in Ref.~\cite{Boeltzig22} were not included in our calculations. Besides, the rates have not been supplemented with Hauser-Feshbach theory at high temperatures to account for unobserved high-energy resonances. However, this only affects the rates above 8 GK.

\section{Conclusion}
In this work, up-to-date $^{23}\text{Na}(p,\gamma)$ direct capture S factor and reaction rates are presented resulting from a $^{23}\text{Na}(^3\text{He},d)$ proton stripping reaction measurement. Spectroscopic factors for many bound states were extracted and used to calculate the direct capture cross section both with the hard-sphere scattering and zero-range scattering approximations. The new rate is smaller than previous studies by $\sim43\%$ at $T_9 < 0.04$ GK, mainly due to the systematic difference between the $C^2S$ values measured from this work and Garrett \textit{et al.}~\cite{Garrett78}. Our results are consistent with the $^{23}\text{Na}(p,\gamma)$ measurement from Ref.~\cite{Boeltzig22}, where they reported an extrapolated nonresonant S factor $\sim50\%$ smaller than that from Garrett's $C^2S$ measurement. 

The present results were obtained with modern global optical models~\cite{Liang09, Daehnick80}, minimized systematic cross section errors from beam and target effects, and a Bayesian MCMC method that rigorously includes all sources of uncertainties. They were rigorously compared with the results of Garrett \textit{et al.}, where we found that model assumptions could not explain the large differences in spectroscopic factors between our results. We confirmed the suggestion from Ref.~\cite{Boeltzig22} that the larger direct capture cross section in previous studies was probably an experimental effect.

Despite these findings, large uncertainties in the direct capture cross section still persist from: (i) the $10.7$ MeV doublet states, which contribute the most but were not sufficiently resolved in our spectra, and (ii) the 1.3 MeV state and 4.2 MeV doublet states that were not included in our measurement. Those can be improved in future work.

 \clearpage
\begin{acknowledgments}
We wish to acknowledge the support of all the TUNL technical staff. Special thanks to Christian Iliadis for his support with the direct capture calculations in this work. \\
This material is based partly upon work supported by the U.S. Department of Energy, Office of Science, Office of Nuclear Physics, under Contract Nos. DE-FG02-97ER41033 and DE-FG02-97ER41042.
This work is supported by the U.S. Department of Energy, Office of Science, Office of Nuclear Physics, under Award Numbers DE-SC0017799 and under Contract No. DE-FG02-97ER41041.

\end{acknowledgments}

\bibliography{apssamp}

\end{document}